\documentclass[prd,aps,floats,floatfix,superscriptaddress,preprintnumbers,
showpacs,eqsecnum,nofootinbib,twocolumn]{revtex4}
\usepackage{latexsym,array,theorem,mathrsfs,bm,float}

\usepackage{psfrag}
\usepackage{amsfonts,amsmath,amssymb,latexsym,array,afterpage,
theorem,mathrsfs,bm,float,epsfig,color,graphicx,tabularx,here,multirow}
\newcommand{\nn}{\nonumber \\}
\newcommand{\bea}{\begin{eqnarray}}
\newcommand{\ena}{\end{eqnarray}}
\newcommand{\beann}{\begin{eqnarray*}}
\newcommand{\enann}{\end{eqnarray*}}

\newcommand{\lsim}{\, \mbox{\raisebox{-1.ex}
{$\stackrel{\textstyle<}{\textstyle\sim}$}}\,}
\newcommand{\ma}[1]{\mbox{$\mathcal{#1}$}}

\newcommand{\ti}{\tilde}

\newcommand{\vect}[1]{\!\!\!\mbox{ \boldmath $#1$}}
\newcommand{\calhR}[1]{\raisebox{2ex}{\tiny ({\em h})}\hspace{-0.8em}{\ma R}}

\begin{document}

\title{
AdS Monopole Black Hole and Phase Transition
}


\author{Shoichiro {\sc Miyashita}}
\email{miyashita-at-gravity.phys.waseda.ac.jp}
\author{Kei-ichi {\sc Maeda}}
\email{maeda-at-waseda.jp}
\address{Department of Physics, Waseda University, 
Okubo 3-4-1, Shinjuku, Tokyo 169-8555, Japan}


\date{\today}

\begin{abstract}
We study the Einstein-SO(3)Yang-Mills-Higgs system with a negative cosmological constant, 
and 
find the monopole black hole solutions as well as the trivial Reissner-Nordstr\"{o}m black hole.
We  discuss thermodynamical stability of the monopole black hole
 in an isolated system.
We expect a phase transition between those two black holes 
 when the mass of a black hole 
increases or decreases. 
The type of phase transition depends on the cosmological constant 
$\Lambda$
as well as the vacuum expectation value $v$ and the coupling constant $\lambda$
of the Higgs field. 
Fixing $\lambda$ small, we find
there are two critical values of the cosmological constant $\Lambda_{\rm cr (1)}(v)$ and 
$\Lambda_{\rm cr(2)}(v)$, which depend on $v$.
If $\Lambda_{\rm cr(1)}(v)<\Lambda (<0)$, we find the first order transition, while 
if $\Lambda_{\rm cr(2)}(v)<\Lambda<\Lambda_{\rm cr(1)}(v)$, the transition becomes
 second order.  For the case of $\Lambda_{b}(v)<\Lambda<\Lambda_{\rm (2)}(v)$,
 we again find the first order irreversible 
 transition from the monopole black hole to the extreme 
  Reissner-Nordstr\"{o}m black hole.
  Beyond $\Lambda_{b}(v)$, no monopole black hole exists.
 We also discuss thermodynamical properties of the monopole black hole
 in a thermal bath system.
\end{abstract}


\maketitle

\section{Introduction}
A black hole (BH) has been a central object in relativistic astrophysics as well as 
in gravitational physics 
since the discovery of the Schwarzschild solution.
 One of the most important findings is the uniqueness of the Kerr solution\cite{uniqueness}.
 An axisymmetric stationary vacuum solution,  which describes a rotating BH,
  is uniquely determined by the Kerr solution. 
 In the Einstein-Maxwell theory, it was also proven that a BH solution is
  uniquely given by the Kerr-Newman solution with its global charges. 
This is the so-called black hole no-hair theorem.

However, if we extend  matter fields to a more general class, 
we find a different type of solutions, which is called a hairy BH.
After Bartnik and McKinnon (BM) discovered a non-trivial soliton solution in 
the Einstein-Yang-Mills (YM) theory,
the colored BH was found in the same system\cite{colored BH 1,colored BH 2}. 
This hairy BH has the same global charges as those of the Schwarzschild BH.
Hence the uniqueness of a black hole solution is no longer applied when we 
consider YM field.
A YM hair, which is not a global charge, appears for such a BH. 
Although it turns out that the colored BH as well as the BM soliton solution are unstable, in 
the extended models,  asymptotically flat  non-Abelian BH solutions 
were intensively studied a few decades ago\cite{non-Abelian BH 1, non-Abelian BH 2}. 
One of the most interesting non-Abelian BHs is a monopole BH, 
which is obtained in the system with an SU(2) (SO(3)) Yang-Mills field and a real triplet Higgs field. 
Before the BM solution, 
the study of the Einstein-SO(3)Yang-Mills-Higgs (EYMH) system was
performed from the viewpoint of generalization of `t Hooft-Polyakov monopole in a curved spacetime 
\cite{monopole in curve}. 
The asymptotically flat monopole BH  was found
\cite{monopole BH 1,monopole BH 2, monopole BH 3} soon after the discovery of a self-gravitating `t Hooft Polyakov monopole,
and the various properties were investigated
\cite{monopole stability,monopole BH 4,monopole BH 5,monopole BH 6, monopole BH 7,monopole BH 8,monopole BH 9,monopole BH 10}.
The monopole BH is magnetically charged BH with a global monopole charge, 
and its event horizon is located inside a gravitating monopole. 
One of the most important properties is its stability. 
In this system, there are two BH solutions, the monopole BH and the Reissner-Nordstr\"{o}m (RN) BH,
 with the same mass. 
  The monopole BH, if it exists, 
  is dynamically stable against the perturbations\cite{monopole stability}.
 While
 the RN BH with the same horizon radius becomes unstable. 
However, since the event horizon of the monopole BH 
exists inside a monopole, which radius is
determined by the coupling constants, when we increase the horizon radius, 
non-trivial monopole structure outside the event horizon disappears, 
resulting in a unique stable RN BH. 
This behavior of stability is well understood by a 
catastrophe theory\cite{catastrophy}.

This phenomenon is also easily reinterpreted via BH thermodynamics\cite{thermodynamics}.
The laws of the BH dynamics look similar to those of thermodynamics. 
This analogy turns to be real physical laws as BH thermodynamics after the discovery of Hawking radiation\cite{Hawking}. 
In BH thermodynamics,  the  ADM mass, the area of the event horizon, and 
the surface gravity correspond to the internal energy, the entropy, and the temperature,  respectively.
The other globally conserved quantities such as the electric (magnetic) charge or the angular momentum are also treated as thermodynamical quantities. 
Using the thermodynamical variables such as BH entropy, we can discuss 
various types of phase transitions depending on the coupling constants and parameters.  

In the SO(3) EYMH case, the monopole BH solution 
is found as a new phase possessing higher entropy than that of the RN BH.
Hence we expect that a phase transition between 
the monopole BH and the RN BH occurs.
Actually it was found the second order phase transition 
 between these two BHs. 
 However, in some parameter region, 
 we also find the ``reversible''  first order phase 
 transition  between these two BHs,
  and the ``non-reversible''  first order phase transition 
 from the monopole BH to the extreme RN BH \cite{monopole BH 4}.

Recently,  in the context of AdS/CFT correspondence
\cite{ADSCFT},  
a black hole with asymptotically anti de Sitter (AdS) spacetime  has been 
attracting the most attention.
Such a black hole has been studied intensively during the last decade \cite{BH in ADS}.
A particle-like solution and  black hole  in the EYM system with a negative cosmological constant 
were found in 1999 and 2000\cite{AdS colored BH,BH soliton}. 
It turns out that such solutions are stable.
Their thermodynamical properties were also analyzed 
\cite{Thermo SU2YM 1,Thermo SU2YM 2}.

Here we study the EYMH system with a negative cosmological constant.
A particle-like solution (a gravitating monopole) was found in \cite{ADS monopole,ADS monopole 2}.
We construct a magnetically charged BH solution and  investigate 
its classical properties and thermodynamimcal properties. 
We examine the dependence of a negative cosmological constant on 
the phase transition. 
We should note that different spacetime solutions in the  SO(3)EYMH system 
were found and some of them were applied it to a holographic analysis for a strongly coupled gauge field
\cite{other EYMH 0, other EYMH 1,other EYMH 2,other EYMH 3,other EYMH 4,other EYMH 5,other EYMH 6,other EYMH 7,other EYMH 8}.

The paper is organized as follows. 
We introduce the Einstein-Yang-Mills-Higgs system
and provide the basic equations for a spherically symmetric static spacetime
in \S \ref{EYMHwncc}.
 In \S \ref{AdSMBH}, we present
a monopole black hole as well as a self-gravitating monopole solution.
 We then analyze the thermodynamics of the present system 
 in \S \ref{THAdSMBH}. 
 We discuss the phase transition  between the monopole BH and the RN BH, 
 which order depends on a cosmological constant as well as the vacuum expectation value 
 (VEV)
 of the Higgs field.
 We also show how the Hawking-Page transition occurs 
 in the thermal equilibrium state.
We summarize our results and give some remarks in \S 
\ref{summary}.


\section{Einstein-Yang-Mills-Higgs system with a 
cosmological constant}
\label{EYMHwncc}

\subsection{Basic equations}

We consider the Einstein-Yang-Mills-Higgs (EYMH) system with a 
cosmological constant $\Lambda$, which action is given by
\begin{equation}
\begin{array}{l}
\displaystyle I_{\rm EH}=\int d^4 x \sqrt{-g}\left[\mathcal{L}_{\rm  G} +\mathcal{L}_{\rm YM} +\mathcal{L}_{\rm  H} \right]\,, \\
\displaystyle \mathcal{L}_{\rm  G} =\frac{1}{16 \pi G}(R-2\Lambda)\, \\
\\
\displaystyle \mathcal{L}_{\rm YM}=-\frac{1}{4}F^{(a)}_{\mu\nu}F^{(a)\mu\nu}\, \\
\\
\displaystyle \mathcal{L}_{\rm H} =-\frac{1}{2}D_{\mu}\Phi^{(a)}
 D^{\mu}\Phi^{(a)} 
-\frac{\lambda}{4}(\Phi^{(a)} \Phi^{(a)} -v^2)^2\, \\
\end{array}
\label{ads0}
\end{equation}
where $F^{(a)}_{\mu\nu}$ is SO(3) YM field defined by 
\begin{equation}
F^{(a)}_{\mu\nu}=\partial_{\mu}A^{(a)}_{\nu}-\partial_{\nu}A^{(a)}_{\mu}
-e\epsilon^{abc}A^{(b)}_{\mu}A^{(c)}_{\nu}\,,
\label{mono2}
\end{equation}
with $A^{(a)}_{\mu}$ and $e$ being 
the gauge field potential and the YM coupling constant.
$\Phi^{(a)}$, $v$ and $\lambda$ are a real Higgs triplet,
its VEV and selfcoupling constant, respectively.
\footnote{The Greek indices ($\mu, \nu, \cdots$) and the Latin indices ($i, j, \cdots$) run 
the spacetime coordinates ($0,1,2,3$) and the spatial coordinates ($1,2,3$), while the Latin indices
 with round brackets ($(a), (b), \cdots$) describe the internal coordinates. } 
 
The covariant derivative $D_{\mu}$ is defined by 
\begin{equation}
D_{\mu}\Phi^{(a)}=\partial_{\mu}\Phi^{(a)}-e\epsilon^{abc}A^{(b)}_{\mu}\Phi^{(c)}.
\label{mono3}
\end{equation}

The Higgs mass $m_{\rm H}$ and gauge boson mass $m_A$, and their Compton wave lengths 
$\ell_{\rm H}$ and $\ell_A$, are given by 
\bea
m_{\rm H}=&\sqrt{\lambda}v\,,~~&\ell_{\rm H}={1\over \sqrt{\lambda}v}
\nonumber \\
m_A=&ev\,,~~&\ell_A={1\over ev}
\,.
\ena

~\\
We will normalize all variables by the Compton wave length of the YM field, 
$\ell_A$, with the units of $\hbar=c=1$.  The 
Planck mass is 
defined by $M_{\rm PL}=G^{-1/2}$.

\subsection{A spherically symmetric static system}
In order to discuss a gravitating monopole and monopole BH,
 we consider a spherically symmetric static system, and 
 assume the Hedgehog ansatz for the SO(3) YM field and Higgs field as
\begin{equation}
\begin{array}{l}
\Phi^{(a)}=v\hat{r}^{(a)}h(r)\,,\\ 
\displaystyle A^{(a)}_{i}=w^{(c)}_{i}\epsilon^{abc}\hat{r}^{(b)}\frac{1-w(r)}{er}
\,,
\end{array}
\label{mono4}
\end{equation}
where $w^{(a)}_{i}$ is the triad of the 3-space
 and $\hat{r}^{(a)}$ is the unit ``radial" vector
 in the internal space.

The static and spherically symmetric metric is described as follows:
\begin{equation}
ds^{2}=-f(r) e^{-2\delta(r)}dt^{2}+f(r)^{-1}dr^2
+r^2 d \theta^2 +r^2 \sin^2 \theta d \phi^2 \,,
\label{ads1}
\end{equation}
where 
\begin{equation}
f(r)=1-\frac{2Gm(r)}{r}-\frac{\Lambda}{3}r^{2}\,,
\end{equation}
and $\delta(r)$ are metric functions, which depend  only on the radial coordinate $r$.

Varying the action (\ref{ads0}) with respect to the metric $g^{\mu\nu}$, the 
Higgs field $\Phi^{(a)}$ and the YM gauge field 
$A_\mu^{(a)}$, we find four basic equations.

Introducing the dimensionless variables $\tilde{r}\equiv r/\ell_A$, 
$\tilde{m}\equiv Gm/\ell_A$ 
as well as the dimensionless parameters, $\tilde{v} \equiv v/M_{\rm PL}$,
 $\tilde{\Lambda} \equiv \Lambda \ell_A^{2}$,
the four basic equations are written as:
\begin{widetext}
\bea
 \frac{d\tilde{m}}{d\tilde{r}}&=&4\pi \tilde{v}^{2}
\left[ f\left\{ \left( \frac{dw}{d\tilde{r}} \right)^{2}+\frac{\tilde{r}^{2}}{2}
\left( \frac{dh}{d\tilde{r}} \right)^{2} \right\}+\frac{(w^{2}-1)^{2}}{2\tilde{r}^{2}}
+w^{2}h^{2} \right]+\pi\bar{\lambda}\tilde{v}^{2}\tilde{r}^{2}(h^{2}-1)^{2}
\label{ads6}
\\
 \frac{d\delta}{d\tilde{r}}&=&-8\pi\tilde{v}^{2}\tilde{r}
\left[ \frac{1}{\tilde{r}^{2}}\left( \frac{dw}{d\tilde{r}} \right)^{2}+\frac{1}{2}
\left( \frac{dh}{d\tilde{r}} \right)^{2} \right]
\label{ads7}
\\
\frac{d^{2}w}{d\tilde{r}^{2}}&=&\frac{1}{\tilde{r}^{2}f}\left[w(w^{2}-1
+\tilde{r}^{2}h^{2})-2\left(\tilde{m}-\frac{\tilde{\Lambda}}{3}\tilde{r}^{3} \right)
\frac{dw}{d\tilde{r}} \right. 
\nn
&& \left.
+8\pi\tilde{v}^{2}\tilde{r}\frac{dw}{d\tilde{r}}\left\{ 
\frac{(w^{2}-1)^{2}}{2\tilde{r}^{2}}+w^{2}h^{2}+{\bar{\lambda}\over 4}
\tilde{r}^{2}(h^{2}-1)^{2} \right\}\right]
\label{ads8}
\\
\frac{d^{2}h}{d\tilde{r}^{2}}&=&-\frac{2}{\tilde{r}}
\frac{dh}{d\tilde{r}}+\frac{1}{\tilde{r}^{2}f}\left[2hw^{2}+\bar{\lambda}
\tilde{r}^{2}h(h^{2}-1)-2\left( \tilde{m}-\frac{\tilde{\Lambda}}{3}\tilde{r}^{3} 
\right)\frac{dh}{d\tilde{r}} \right.
\nn
&& \left.
+8\pi\tilde{v}^{2}\tilde{r}\frac{dh}{d\tilde{r}}\left\{ 
\frac{(w^{2}-1)^{2}}{2\tilde{r}^{2}}+w^{2}h^{2}+{\bar{\lambda}\over 4}
\tilde{r}^{2}(h^{2}-1)^{2} \right\}\right]
\,,
\label{ads9}
\ena
where we define $\bar{\lambda}=\lambda/e^2$.
\end{widetext}

\subsection{Boundary conditions}
Next, we consider the boundary conditions.
\subsubsection{boundary conditions at a center or at a horizon} 
For a gravitating monopole, we  impose a regularity at the center ($\tilde r=0$).
We find
\bea
w(\tilde{r})&=&1-c_{w}\tilde{r}^{2}+ o(\tilde{r}^{3})
\\
h(\tilde{r})&=& c_{h}\tilde{r}+o(\tilde{r}^{2})
\\
\tilde{m}(\tilde{r})&=&  c_m \tilde{r}^{3} + o(\tilde{r}^{4})
\ena
with 
\bea
c_m=\frac{4\pi}{3}\tilde{v}^{2}\left( 6c_{w}^{2} + \frac{3}{2}c_{h}^{2}
+\frac{1}{4}\bar{\lambda}  \right)
\,.
\ena
The constants $c_w$ and $c_h$ are the shooting parameters,
which are fixed by the asymptotically AdS condition.

For a black hole solution, we impose a regularity at the horizon $\tilde r=\tilde r_H$, 
which is defined by 
$
f(\tilde r_H)=0\,,
$ i.e.,
\bea
 \tilde{m}(\tilde{r}_{H})=\frac{\tilde{r}_{H}}{2} 
\left(1-\frac{\tilde{\Lambda}}{3}\tilde{r}_{H}^{2}  \right)
\,.
\label{ads13}
\ena

\begin{widetext}
We then find 
the first derivatives of $w(\tilde{r})$ and $h(\tilde{r})$ on the horizon 
are given as
\bea
\displaystyle \left. \frac{dw}{d\tilde{r}}\right|_{\tilde{r}=\tilde{r}_{H}}=
\frac{\tilde{r}_{H} w_{H}(w_{H}^{2}-1+\tilde{r}_{H}^{2}h_{H}^{2})}{
(1-\tilde{\Lambda}\tilde{r}_{H}^{2})-2\pi\tilde{v}^{2}
\left[ 2(w_{H}^{2}-1)^{2}+4\tilde{r}_{H}^{2}w_{H}^{2}h_{H}^{2}
+\bar{\lambda}\tilde{r}_{H}^{4}(h_{H}^{2}-1)^{2} \right] }
\label{ads10}
\\
\displaystyle \left. \frac{dh}{d\tilde{r}}\right|_{\tilde{r}=\tilde{r}_{H}}=
\frac{\tilde{r}_{H}h_{H}(2w_{H}^{2}+\bar{\lambda}\tilde{r}_{H}^{2}(h_{H}^{2}-1))}{
(1-\tilde{\Lambda}\tilde{r}_{H}^{2})-2\pi\tilde{v}^{2}
\left[ 2(w_{H}^{2}-1)^{2}+4\tilde{r}_{H}^{2}w_{H}^{2}h_{H}^{2}
+\bar{\lambda}\tilde{r}_{H}^{4}(h_{H}^{2}-1)^{2} \right] }
\,,
\label{ads11}
\ena
\end{widetext}
where 
$w_H=w(\tilde{r}_{H})$ and $h_H=h(\tilde{r}_{H})$. 
Giving  the horizon radius $ \ti{r}_H$,
we have to fix the remaining two constants ($w_H$ and $h_H$) by 
the shooting method to find a regular BH with an asymptotically AdS spacetime.
We assume that 
$f(\tilde{r})$ is always positive outside the horizon $\tilde r_H$ 
in order not to make an additional horizon.

\subsubsection{boundary conditions at spatial infinity}
We assume that a gravitating monopole or a monopole black hole approaches
an AdS spacetime at spatial infinity.
 Then we assume that the mass function 
$\tilde{m}$ is finite at infinity, i.e.,
\bea
\tilde{m}(\infty)={\rm constant}
\,,
\label{ads15}
\ena
which implies 
\bea
w(\infty)=0\,,
~~
h(\infty)=1\,, 
\label{ads17}
\ena
as well as 
\begin{equation}
\delta(\infty)=0\,.
\label{ads12}
\end{equation}

Under the above boundary conditions, we solve the basic equations.
Before giving a non-trivial gravitating monopole or a monopole BH solution,
we just show a trivial RN BH solution as follows:
\subsection*{Trivial Solution}
When we assume the variables as
\begin{equation}
w(r)=0, \, h(r)=1,~~ {\rm and }~~ \delta(r)=0\,,
\end{equation}
Eqs. (\ref{ads7}), (\ref{ads8}) and (\ref{ads9}) are trivially satisfied, and 
Eq. (\ref{ads6}) becomes
\begin{equation}
m'(r)=\frac{2\pi}{G e^{2}r^{2}}
\,,
\end{equation}
which is easily integrated as
\begin{equation}
m(r)=M-\frac{2\pi}{G e^{2}r},
\end{equation} 
where
\begin{equation}
M=\frac{r_{H}}{2G}\left( 1+\frac{4\pi}{e^{2}r_{H}^{2}}-\frac{\Lambda}{3}r_{H}^{2} \right).
\end{equation}
This is the Reissner-Nordstr\"{o}m AdS BH solution with the ADM mass $M$ and 
the magnetic charge $2\pi^{1/2}/e$.

\section{Non-trivial Solutions}
\label{AdSMBH}

As mentioned before, we have performed shooting for two parameters numerically in order to 
obtain non trivial solutions; both a self-gravitating monopole  and a monopole BH. 
Although the monopole solution was obtained in \cite{ADS monopole,ADS monopole 2}, we show 
both solutions below.

\subsection{Self-gravitating Monopole}
\begin{figure}
	\includegraphics[width=6cm]{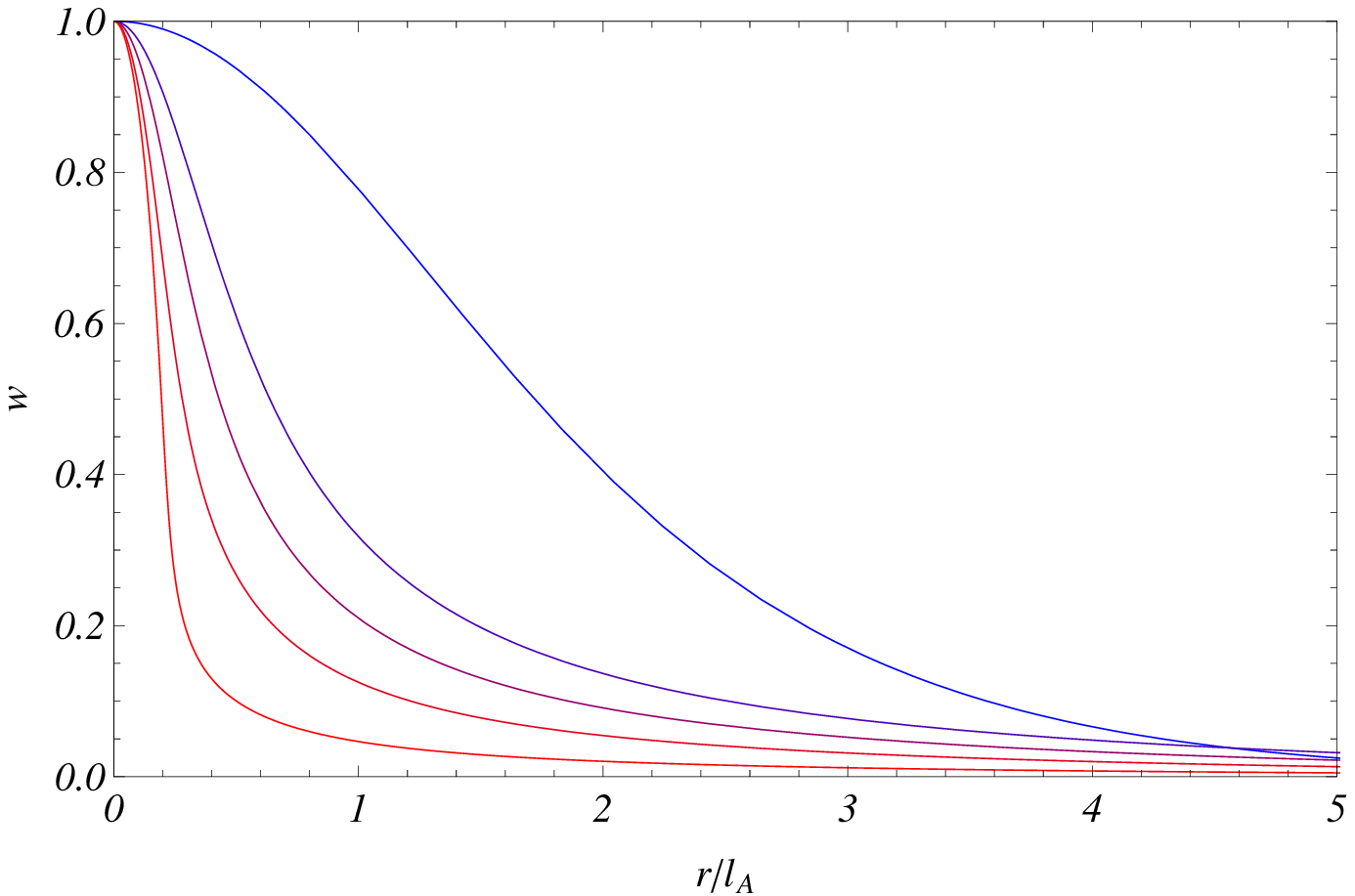}
	\includegraphics[width=6cm]{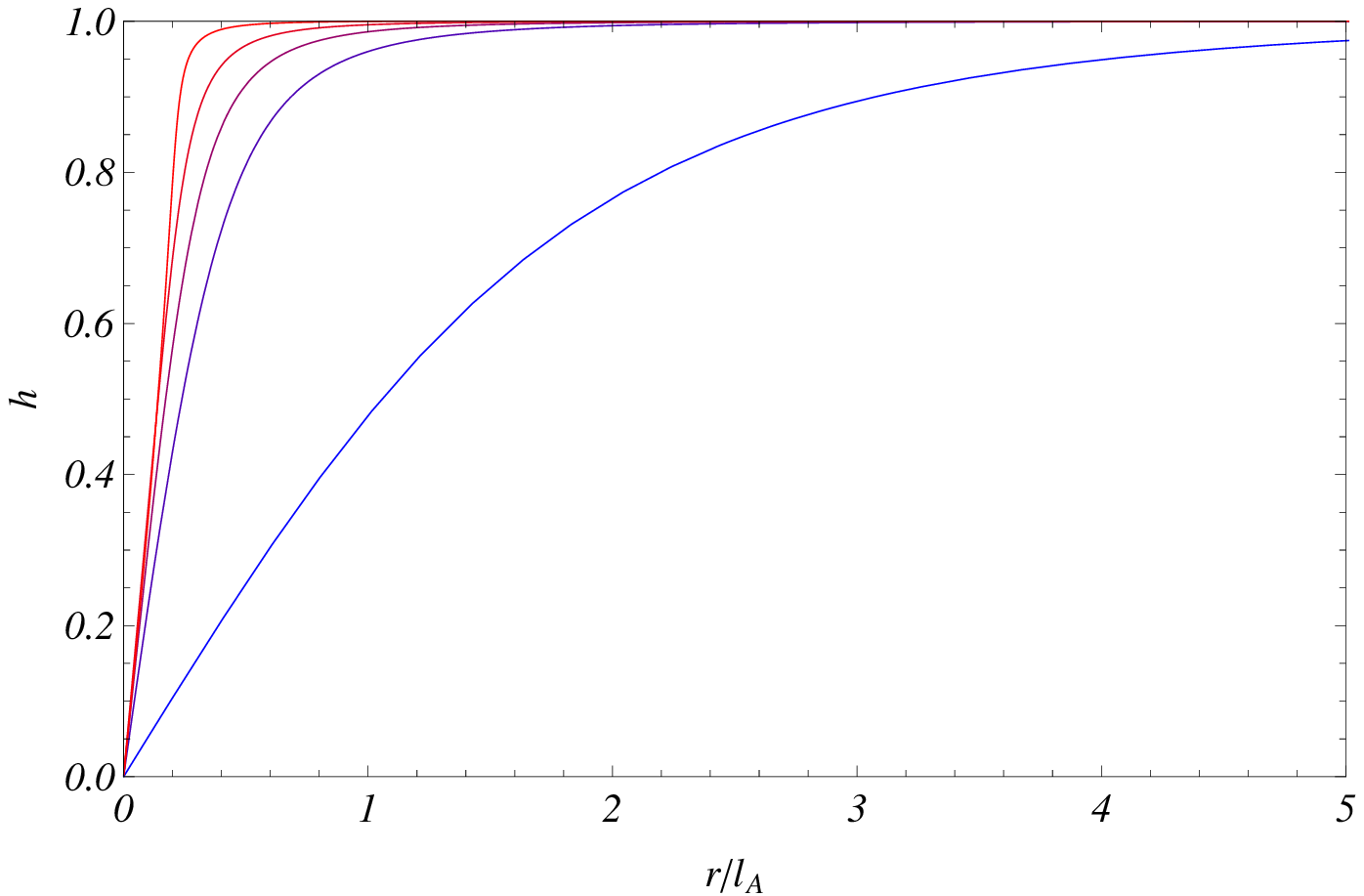}
	\caption{A self-gravitating monopole. We show the Higgs field $h$ and the gauge potential $w$ in terms of the radius $\ti{r}$. We set $\bar{\lambda}=0.1$, and $\ti{v}=0.1$. We choose $\ti{\Lambda}=0, -10, -20, -30$ and $\ti{\Lambda}=-33.8\approx\ti{\Lambda}_{\rm b}$, 
from the top curve to the bottom one in the case of $w$,  and 
from the bottom one to the top one for $h$.}
\label{fig-monopole1}
\end{figure}

We show some examples of gravitating monopole solutions in Fig.\ref{fig-monopole1}
for the case of $\bar{\lambda}=0.1$ and $\tilde{v}=0.1$.
We choose $\tilde{\Lambda}=-1, -10, -20, -30$, and $-33.8 \approx\tilde{\Lambda}_{\rm b}$,
beyond which there exists no monopole solution.

\begin{figure}
	\includegraphics[width=6cm]{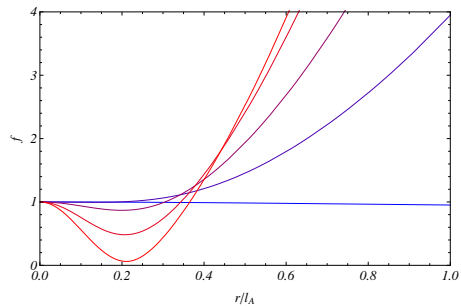}
	\caption{The behavior of $f(\ti{r})$ of the monopole solution, $\ti{v}=0.1$ and  $\bar{\lambda}=0.1$. The curves denote for the cases of  $\ti{\Lambda}=0,-10,-20,-30$ and
	 $-33.8\approx\ti{\Lambda}_{\rm b}$ from the top.}
\label{fig-monopole2}
\end{figure}

As $|\tilde \Lambda|$ increases, the monopole radius shrinks 
in the unit of $\ell_{A}=1$, and
 as shown in Fig.\ref{fig-monopole2}, 
the minimum of the metric function $f(r)$ decreases 
and eventually vanishes at some radius ($r\approx 0.21 \ell_A$)
when the cosmological constant reaches the boundary value 
$\tilde{\Lambda}_{\rm b}(\tilde v)$,
which depend on $\tilde v$.\footnote[2]{
The boundary value also depends on $\bar \lambda$, but 
we fix the value of $\bar \lambda$ in the text 
 ($\bar{\lambda}=0.1$) because it shows a most variety of phase behaviors.
See Appendix A about its dependence for the case of $\tilde \Lambda=0$.
We also find the similar behavior for the case of $\tilde \Lambda<0$.
}
It means that if the typical monopole scale ($\sim \ell_A$) is much 
larger than the AdS curvature length 
$\ell_{\Lambda}=\sqrt{-3/\Lambda}$, no monopole solution is possible.
Beyond the critical value, we just find the RN AdS BH.

\begin{figure}
   	\includegraphics[width=6cm]{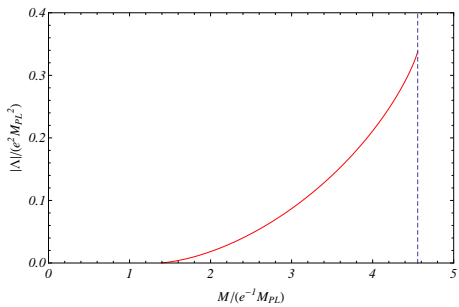}
	\caption{The relation between the monopole mass $M$
	 and the cosmological constant  $\Lambda$. The dashed line 
	 ($M\approx4.55M_{\rm PL}/e$) represents the mass of extreme RN AdS BH when $\ti{\Lambda}=
	 -33.8 \approx \ti{\Lambda}_{\rm b}$. We set $\bar{\lambda}=0.1$ and $\ti{v}=0.1$ }
\label{fig-monopole4}
\end{figure}

We also show the monopole mass $M$ in terms of the cosmological constant 
$\tilde \Lambda$ in Fig. \ref{fig-monopole4}.
The mass is determined by the cosmological constant $\tilde \Lambda$ 
as well as the coupling constants $\bar \lambda$ and the VEV $\tilde v$. 
We confirm that 
the mass $M$ at the boundary value of the cosmological constant 
 is the same as that of the extreme RN AdS BH.

\begin{figure}[h]
	\includegraphics[width=6cm]{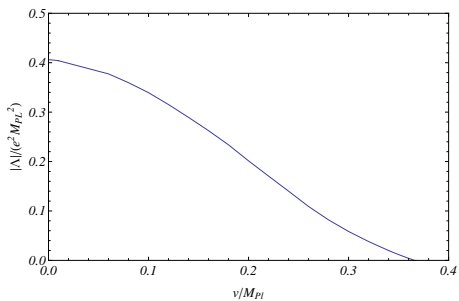}
	\caption{The  boundary value of the cosmological constant 
	($\Lambda=\Lambda_{\rm b}(v)$)
	in the $v$-$\Lambda$ plane,
	beyond which  no monopole solution can exist.
	We set $\bar \lambda=0.1$}
\label{fig-monopole3}
\end{figure}

We know the similar behavior in the case of the asymptotically flat spacetime
(See Appendix A).
There exists a critical value of $\tilde v$, beyond which there is no monopole
and the RN BH is obtained.
When we include a cosmological constant, 
since the monopole structure is determined by  $\tilde \Lambda$ as well as $\ti{v}$, 
there is a boundary value of the cosmological constant, 
\bea
\tilde \Lambda=\tilde \Lambda_{\rm b}(\tilde v)\,,
\ena
beyond which no monopole solution can exist.
We show this boundary curve in the $v$-$\Lambda$ plane 
in Fig. \ref{fig-monopole3}.
We obtain the monopole solution as well as the RN AdS BH  inside the boundary. 
While, we find only  the  RN AdS BH solution outside of this boundary.

\subsection{Monopole Black Hole}

\begin{figure}[h]
\includegraphics[width=6cm]{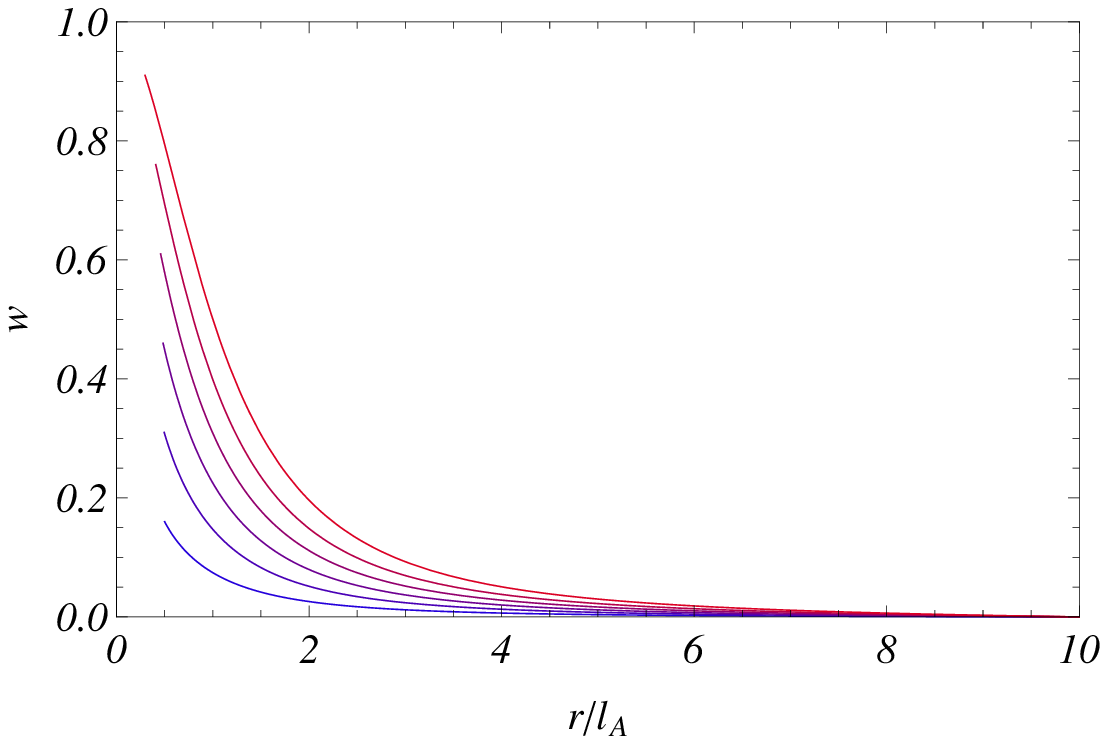}
\\
(a)
\\[1em]
\hskip -.5cm
\includegraphics[width=6cm]{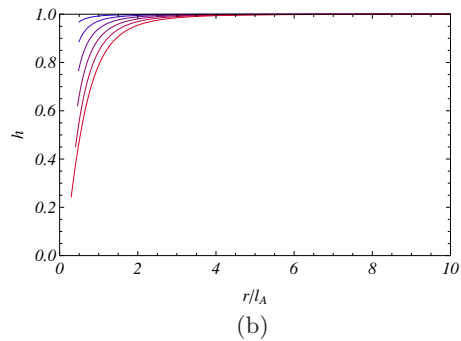}
\\
(b)
\\[1em]
\hskip -.5cm
\includegraphics[width=6cm]{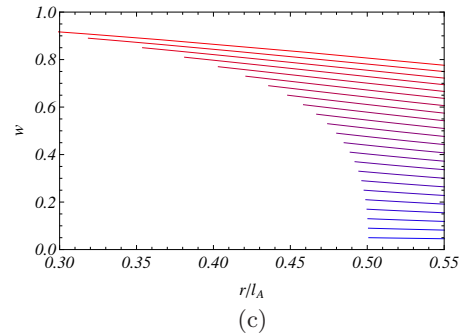}
\\
(c)
\caption{(a) The gauge potential  $w(\tilde{r})$ 
and  in terms of the radius $\tilde r$.
We set  $\bar{\lambda}=0.1$, $\tilde{v}=0.1$, and $\tilde{\Lambda}=-1$.
We choose $w_{H}=0.01, 0.16, 0.31, 046, 061, 0.76$, and $w_{H}=0.91$ 
 from the bottom curve to the top one, respectively. 
 (b) The Higgs field $h(\tilde{r})$ with the same values of $w_{H}$
  from the top curve to the bottom one, respectively.
 In the limit of $r_H\rightarrow 0$, $w_H$ increases while 
 $h_H$ decreases. The solution approaches to the gravitating monopole.
 We also show the near horizon behavior of $w(\tilde{r})$ in (c). 
 The curves correspond to the boundary values of  $w_{H}=0.01 \sim 0.93$ 
 with the interval of 0.04.
 $w_{H}$ decreases as the horizon radius gets larger.}
\label{fig-ads1}
\end{figure}

\begin{figure}[h]
\includegraphics[width=6cm]{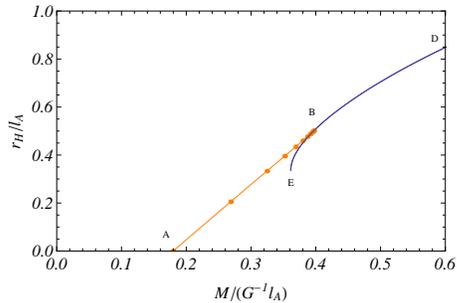}
\caption{The $\tilde M$-$\tilde r_H$ relation for the case of $\ti{\Lambda}=-1$. 
The blue curve denotes the RN AdS BH.}
\label{fig-M-rH1}
\end{figure}

Now we present a monopole BH solution. 
In Fig. \ref{fig-ads1}, we 
depict the gauge potential $w(\ti{r})$ and the Higgs field $h(\ti{r})$.
We set  $\bar{\lambda}=0.1$, $\tilde{v}=0.1$, and $\tilde{\Lambda}=-1$.
 In the limit of $r_H\rightarrow 0$, $w_H$ increases while 
 $h_H$ decreases. The solution approaches to the gravitating monopole.
The qualitative behaviors of $w(\tilde{r})$ and $h(\tilde{r})$ are similar to 
those of the monopole BH without a cosmological constant, 
which are summarized in Appendix A.
As  $w_H$ decreases,
 the gauge potential $w(r)$ drops faster to zero and 
 the Higgs field $h(r)$ increases more rapidly to unity. 
 It means that the size of the monopole structure gets smaller
 because the black hole swallows some portion of the fields.
We also show the near horizon behavior of the gauge potential $w(\tilde{r})$ in Fig. \ref{fig-ads1} (c).
As  $w_H$ decreases,  the horizon radius $r_H$ increases.
 We find that there exists a maximum radius of the horizon in the limit of $w_H
 \rightarrow 0$.
 We  show the $\tilde M$-$\tilde r_H$ relations 
 in Fig. \ref{fig-M-rH1}.
The monopole BH solution reaches 
the maximum horizon radius, where 
the monopole BH branch connects with the RN AdS BH one.

\begin{figure}[h]
\includegraphics[width=6cm]{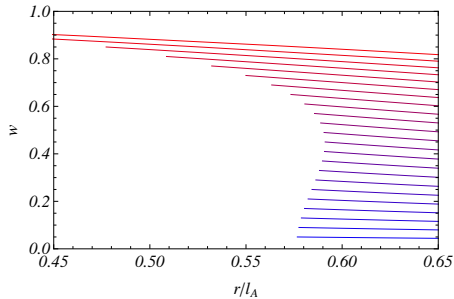}
\caption{The near horizon behavior of the gauge potential $w(\tilde{r})$.
We set $\bar{\lambda}=0.1$ and $\tilde{v}=0.1$
for $\tilde{\Lambda}=-0.1$.
 The curves correspond to the boundary values of  $w_{H}=0.01 \sim 0.93$ 
 with the interval of 0.04. It shows more complicated relation between the boundary values and the horizon radii.}
\label{fig-ads33}
\end{figure}

\begin{figure}[h]
\includegraphics[width=6cm]{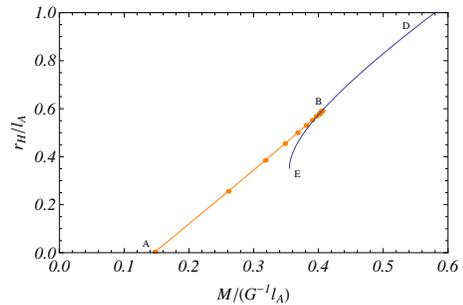}
	\caption{The $\tilde M$-$\tilde r_H$ relation for the case of $\ti{\Lambda}=-0.1$. 
}
\label{fig-M-rH2}
\end{figure}

For the other value of  a cosmological constant,
we may find  the different behavior. 
 In Fig. \ref{fig-ads33}, 
we show the near horizon behavior of $w(\tilde{r})$
  for the case of  $\tilde \Lambda=-0.1$.
As $w_H$ decreases, $r_H$ increases and reaches a maximum at a finite value of $w_H=0.447$,
 and then decreases again.
 In the range of the horizon radius of $0.576\ell_A <r_H<0.591\ell_A$, 
 there are two solutions with different values of $w_H$.
 We  show the $\tilde M$-$\tilde r_H$ relations 
 in Fig.  \ref{fig-M-rH2}.

\begin{figure}[h]
	\includegraphics[width=6cm]{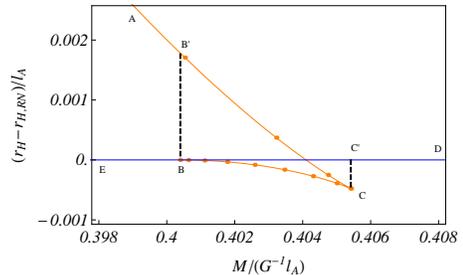}
	\caption{The enlarged figure of Fig. \ref{fig-M-rH2} 
	($\ti{\lambda}=0.1$, $\tilde{v}=0.1$ and $\ti{\Lambda}=-0.1$) near the junction. 
	The vertical axis gives the difference of the horizon radius of the monopole BH 
	 from the RN AdS BH one. The blue curve denotes the RN AdS BH. 
	 When the stable RN AdS BH with larger mass
	 reduces its mass via the Hawking radiation, the evolutionary path  goes 
	  from D $\to$ B and then jumps up  to  B$' \to$ A. 
	  While, if the stable monopole BH with smaller mass
	   increases its mass via the mass accretion, the evolutionary 
	   path goes A $\to$ C and then jumps 
	   up to C$' \to$ D. These are the first order phase transitions between two BHs.}
\label{fig-ads11-2}
\end{figure}

Since the behavior of 
the monopole BH branch is more complicated in this case,
 we enlarge near the junction point in order to see the detail.
We show the difference between the radii of the monopole BH and RN AdS BH
in Fig. \ref{fig-ads11-2}. 
We find there are three BH solutions with the same mass in the mass range of 
$0.4004\ell_A<GM<0.4054\ell_A$, 
which breaks the BH uniqueness.
We find a cusp structure, which we may apply a catastrophe theory to understand
the BH stability just as the case without a cosmological constant\cite{monopole BH 4}.

From the above two examples, we 
understand that there exists a critical value of the cosmological constant 
$\tilde \Lambda=\tilde \Lambda_{\rm cr (1)}(\tilde v)$,
which depends on $\tilde v$, beyond which 
$\tilde M$-$\tilde r_H$ relation is monotonic, i.e., 
only one monopole exists for a given mass $M$.
In the asymptotically flat case ($\ti{\Lambda}=0$), 
we find the similar behavior when we change the value of $\tilde v$, i.e.,
$\tilde M$-$\tilde r_H$ relation is monotonic if  $\tilde v$ is smaller
than the critical value, otherwise there exist two monopole BH solutions 
and $\tilde M$-$\tilde r_H$ relation shows a cusp structure.
(See the detail in Appendix A). 
In the asymptotically AdS case, 
we have the  critical value of $\ti{\Lambda}_{\rm cr(1)}(\tilde{v})$
(or the critical curve in the $v$-$\Lambda$ plane).

\begin{figure}[h]
	\includegraphics[width=6cm]{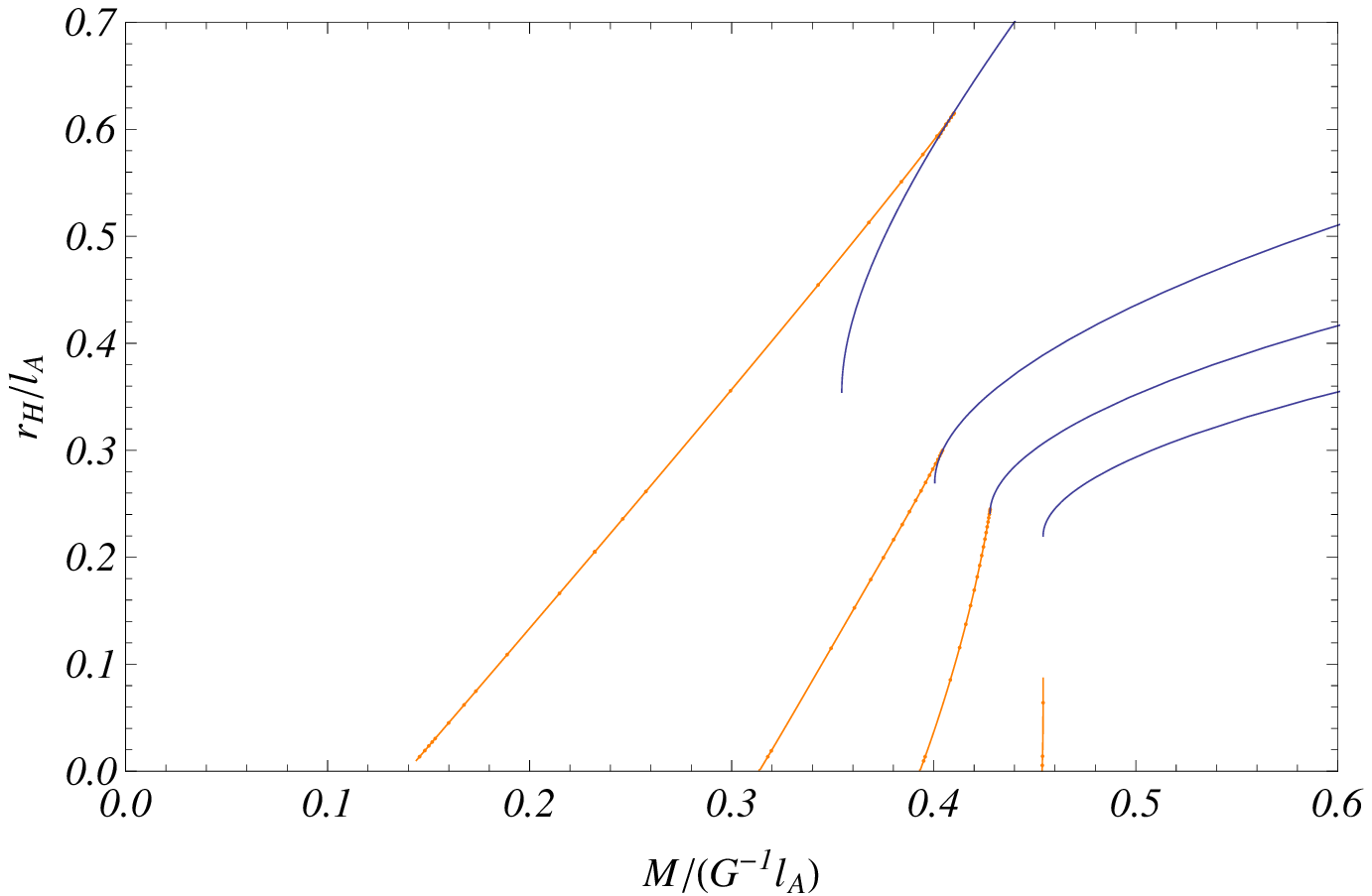}
	\\
	(a)
	\\[1em]
	\includegraphics[width=6cm]{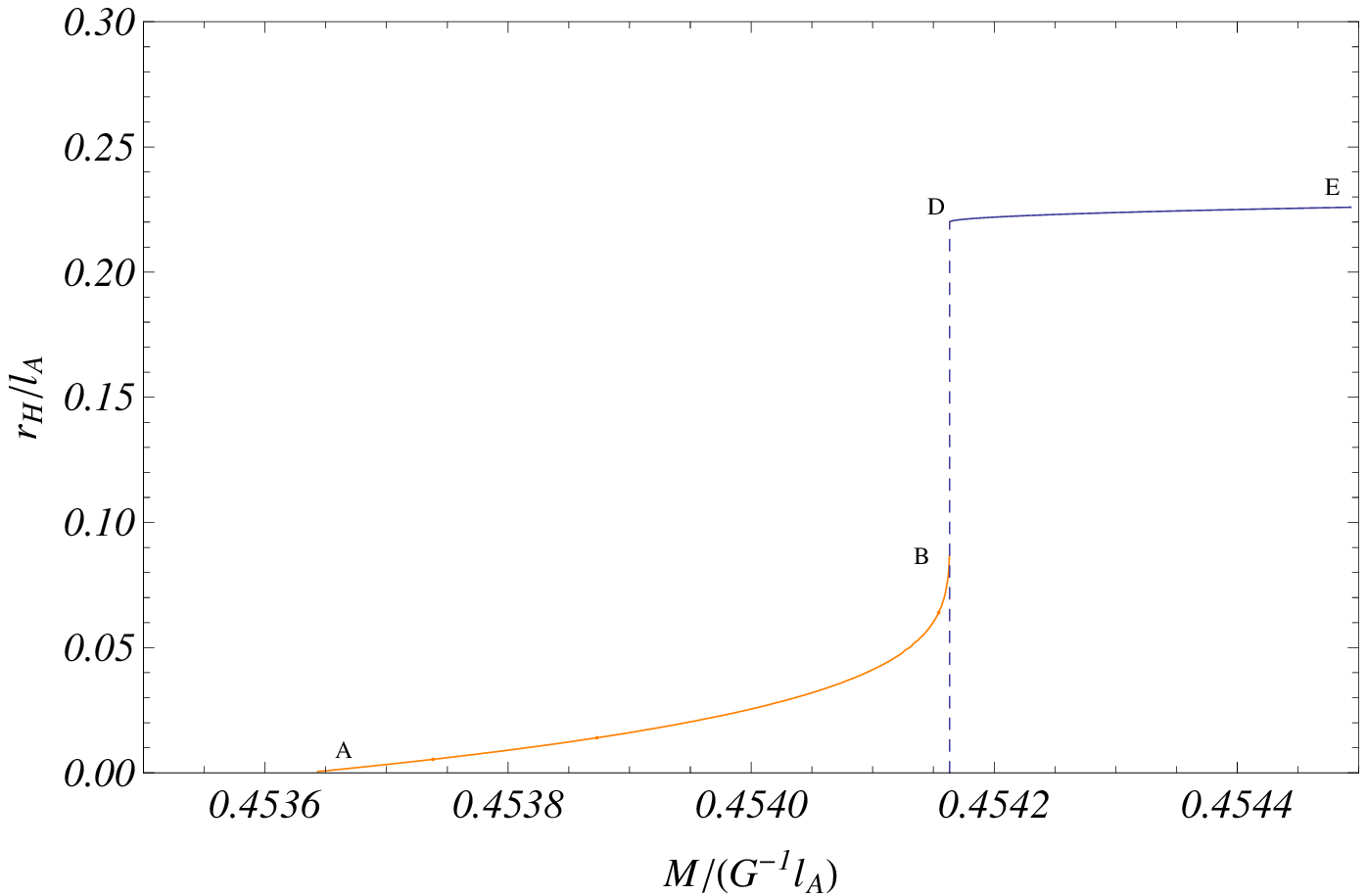}
	\\
	(b)
	\\
	\caption{(a) $M$-$r_{H}$ relation in the case of  $\bar{\lambda}=0.1$ and $\tilde{v}=0.1$.
	We choose $\tilde{\Lambda}=0,-10,-20$ and $-33$. 
		The blue and the orange branches  represents the RN AdS BH and the AdS monopole BH. We find 
	$\tilde{\Lambda}_{\rm cr (1)}\approx -0.68 $, $\tilde{\Lambda}_{\rm cr (2)} \sim -31 $ and 
	$\tilde{\Lambda}_{\rm b}\approx -33.66$.
	 (b) The enlarged figure of $M$-$r_{H}$ relations for the case of  $\tilde{\Lambda}=-33$. The dashed line represents the mass of extreme RN AdS BH.}
\label{fig-ads11}
\end{figure}

We show the $M$-$r_H$ relation for the various values of $\tilde \Lambda$ in Fig. 
\ref{fig-ads11}.
For smaller values of $|\tilde \Lambda|$, we find three BH solutions near the junction point.
When  $|\tilde \Lambda|$ increases beyond 
$|\tilde \Lambda_{\rm cr(1)}|$, the horizon radius of the monopole BH increases monotonically 
and then connects with the RN AdS BH branch.
As $|\tilde \Lambda|$ increases further, the junction point disappears at 
the second critical value 
of the cosmological constant $\tilde \Lambda_{\rm cr(2)}$.
There exists a gap between the horizon radii of the monopole BH and of the RN AdS BH
for the cosmological constant of 
$|\tilde \Lambda|>|\tilde \Lambda_{\rm cr(2)}|$.

\begin{figure}[h]
	\includegraphics[width=6cm]{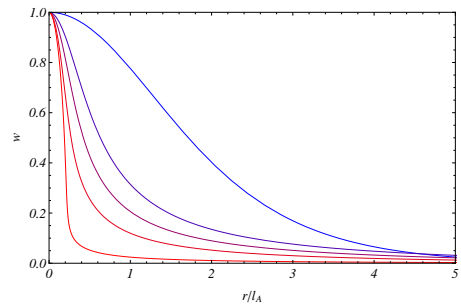}
	\caption{The behavior of $w(\tilde{r})$ fixing $\tilde{r}_{H}=0.01$, $\tilde{v}=0.1$ and $\bar{\lambda}=0.1$ but varying $\tilde{\Lambda}$. $ \tilde{\Lambda}=0,-10,-20,-30$ and 
	$\tilde \Lambda_{\rm b}\approx -33.66$  from the top. }
\label{fig-ads9}
\end{figure}

\begin{figure}[h]
	\includegraphics[width=6cm]{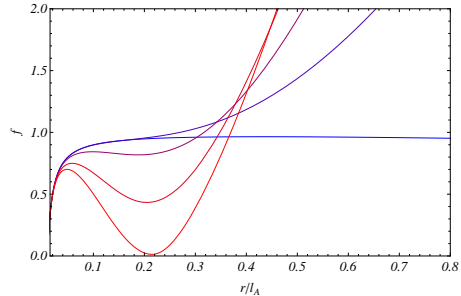}
	\caption{The behavior of $f(\tilde{r})$ fixing $\tilde{r}_{H}=0.01$, $\tilde{v}=0.1$ and $\bar{\lambda}=0.1$ but varying $\tilde{\Lambda}$. $ \tilde{\Lambda}=0,-10,-20,-30$ and 
	$\tilde \Lambda_{\rm b}\approx -33.66$  from the top. }
\label{fig-ads10}
\end{figure}

If $|\tilde \Lambda|$ gets larger furthermore, 
there is the boundary value of the cosmological constant, beyond which 
no monopole exists. So we expect that no monopole BH exists either.
Fig.\ref{fig-ads9} depicts  $w(\tilde{r})$ for the different values of $\tilde \Lambda$ 
with the same horizon radius $\tilde{r}_{H}$, 
$\bar{\lambda}$ and $\tilde{v}$. 
We find that $w(\tilde{r})$ drops  faster to zero as  $|\ti{\Lambda}|$ increases.
This shows that the radius of monopole structure gets smaller by the existence of a 
negative cosmological constant, which we have found in the gravitating-monopole too.
Fig. \ref{fig-ads10} shows
 the behavior of $f(\tilde{r})$ in the case of Fig. \ref{fig-ads9}. 
As $|\tilde{\Lambda}|$ gets larger, the minimum decreases, and eventually reaches zero,
resulting in an extreme RN BH. 
Then monopole BH solution and gravitating monopole solution disappear beyond the boundary value $\tilde \Lambda_{\rm b}$.

We conclude that the effect of $\tilde{\Lambda}$ is the similar to that of $\tilde{v}$ 
as claimed in the work on the AdS monopole\cite{ADS monopole,ADS monopole 2}.  
and
there exist the boundary value $\tilde \Lambda_{\rm b}$, 
beyond which there is no monopole BH solution, as well as 
two critical values $\tilde \Lambda_{\rm cr(1)}$ and $\tilde \Lambda_{\rm cr(2)}$.
These feature is important in the BH thermodynamics and the 
phase transition, which we will discuss in the next section.

\section{Black Hole Thermodynamics and Phase Transition}
\label{THAdSMBH}
As well known, the dynamics of a black hole obeys the thermodynamical laws\cite{thermodynamics}.
Its dynamical stability may be closely related to the thermodynamical stability,
and it may be understood by a catastrophe theory for hairy black holes\cite{monopole BH 4}.
Hence we analyze the thermodynamical variables and discuss the 
thermodynamical stability and possible phase transitions.

First we define the thermodynamical variables.
BH entropy is given by Bekenstein-Hawking area-entropy relation:
\begin{equation}
S=\frac{{\cal A}_H}{4G}=\frac{\pi r_{H}^{2}}{G}
\,,
\label{mono22}
\end{equation}
where ${\cal A}_H=4\pi r_H^2$ is the area of the event horizon.
For convenience of our numerical calculation, we will use the following  
normalized dimensionless entropy:
\begin{equation}
\tilde{S}=e^{2}S=\frac{\pi \tilde{r}_{H}^{2}}{\tilde{v}^{2}}.
\label{mono23}
\end{equation}
The Hawking temperature is given by
\begin{equation}
T=\frac{f'(r_H)}{4\pi}e^{-\delta(r_{H})}
\,,
\label{mono24}
\end{equation}
and the dimensionless temperature $\tilde{T}$ is defined by
\begin{equation}
\left. \tilde{T} \equiv \frac{\sqrt{G}}{e}T=\frac{\tilde{v}}{4\pi}\frac{d f(\tilde{r})}{d \tilde{r}}
\right|_{\tilde{r}=\tilde{r}_{H}} e^{-\delta(\ti{r}_H)}\,.
\end{equation}

In the following subsections, we shall discuss the thermodynamical properties 
of an isolated system and those 
in thermal heat bath, separately.

\subsubsection{Isolated System}
First we consider an isolated system.
The BH entropy is an indicator for the thermal instability.
According to the second law of thermodynamics,
the entropy in an  isolated system never decreases and 
a maximum entropy state is realized. 
In the BH thermodynamics, 
when there exist several BH solutions with the same mass, 
which corresponds to an internal energy in thermodynamics, 
the BH with maximum entropy is thermodynamically stable. 
However, it does not mean that the other BH is always unstable. 
 Although it is not globally stable, it can be locally stable. 
This local thermodynamical stability may correspond to the 
 dynamical stability of BHs.
 Although there is no proof, 
 no counter example is known.

Since the BH entropy is proportional to the horizon area
($\tilde {\cal A}_H=4\pi \tilde r_H^2$), 
the $\tilde{M}$-$\tilde{r}_{H}$ relation is sufficient 
to analyze the thermodynamical stability.
A catastrophe theory may also provide a good tool to judge the stability
\cite{monopole BH 4}.

Classifying the value of the cosmological constant into the following three cases, 
we shall discuss the behavior of the horizon radius and BH phase transition separately.
\\[1em]
(1) \underline{$\tilde \Lambda_{\rm cr (1)}(\tilde v)<\tilde \Lambda (\leq 0)
$}\\
There are two monopole BH solutions  as well as RN AdS BH.
We find a cusp structure in the $\tilde M$-$\tilde r_H$ relation.

In Fig \ref{fig-ads11-2}, we show the $\tilde M$-$\tilde r_H$ relation for the case of 
$\bar \lambda=0.1, \tilde v=0.1$ and $\tilde \Lambda=-0.1$
(Note that $\tilde \Lambda_{\rm cr (1)}\approx -0.68$).
 The RN AdS BH and the monopole BH are stable on the intervals of BD and of AC, 
respectively.
      When the stable RN AdS BH with larger mass
	 reduces its mass via the Hawking radiation, the evolutionary path  goes 
	  from D $\to$ B and then jumps up  to  B$' \to$ A. 
	  While, if the stable monopole BH with smaller mass
	   increases its mass via the mass accretion, the evolutionary 
	   path goes A $\to$ C and then jumps 
	   up to C$' \to$ D. These are the first order phase transitions between two BHs.
\\[.5em]
(2) \underline{$\tilde \Lambda_{\rm cr(2)}(\tilde v)<\tilde \Lambda<
\tilde \Lambda_{\rm cr (1)}(\tilde v)$}\\
There is one monopole BH solution as well as RN AdS BH.
The monopole BH branch connects with the RN AdS BH branch.
This is no cusp structure in the $\tilde M$-$\tilde r_H$ relation
as shown in Fig. \ref{fig-M-rH1}, for which we have chosen 
$\bar \lambda=0.1, \tilde v=0.1$ and $\tilde \Lambda=-1$
(Note that $\tilde \Lambda_{\rm cr(2)}\sim -31$).
The RN AdS BH is unstable in the intervals of BE.
As a result, the evolutionary path when the mass changes 
is either A$\to$B$\to$D or the reverse.
This is the second order phase transition between two BHs.
\\[.5em]
(3) \underline{$\tilde \Lambda_{\rm b}(\tilde v)<
\tilde \Lambda<\tilde \Lambda_{\rm cr (2)}(\tilde v)$}\\
There is one monopole BH solution as well as  RN AdS BH, but those two branches 
are disconnected as shown in Fig. \ref{fig-ads11} (b), for which we have chosen 
$\bar \lambda=0.1, \tilde v=0.1$ and $\tilde \Lambda=-33$
(Note that $\tilde \Lambda_{\rm b}\approx -33.66$). There exists a radius gap at the extreme states between two branches. The evolutionary path when the mass increases goes 
from A$\to$B and then jumps up to D$\to$ E.
It may be the non-reversible first order phase transition.
\\[.5em]
(4) $\tilde \Lambda<\tilde \Lambda_{\rm b}(\tilde v)$
\\
There is no monopole BH solution. Only a stable RN AdS BH exists.
No phase transition occurs.

In order to see the effect of $\ti{\Lambda}$ on the phase transition strength, 
we analyze the magnitude of entropy discontinuity. 
Comparing the $\ti{M}$-$\ti{r}_{H}$ relations in  Fig. \ref{fig-ads11-2} 
($\ti{\Lambda}=-0.1$) with Fig. \ref{fig-mono7} ($\ti{\Lambda}=0$),
 we find the cusp becomes smaller as $|\ti{\Lambda}|$ gets bigger.

\begin{figure}[h]
	\includegraphics[width=6cm]{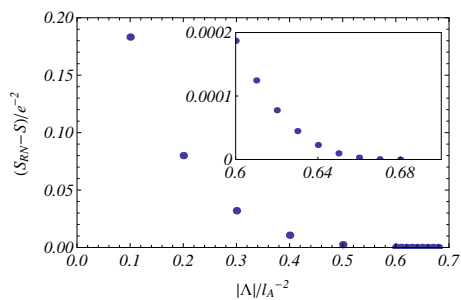}
	\caption{$\tilde{\Lambda}$ dependence of the discontinuity of the BH entropy between 
	the monopole BH and RN (AdS) BH at the cusp point. 
	We set $\bar{\lambda}=0.1$ and 
	$\tilde{v}=0.1$, for which we find $\tilde \Lambda_{\rm cr (1)}\approx -0.68$.
	 The inside figure is the enlarged one near the transition point. }
\label{fig-ads12}
\end{figure}

\begin{figure}[h]
	\includegraphics[width=6cm]{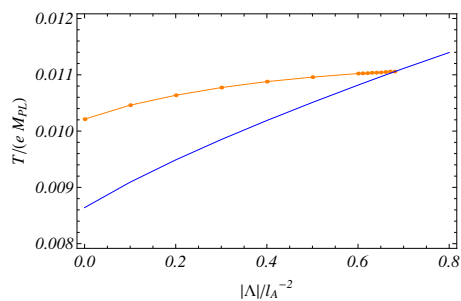}
	\caption{The BH temperatures with respect to the $\tilde{\Lambda}$.
	We set $\bar{\lambda}=0.1$ and $\tilde{v}=0.1$.
	 The orange  branch and the blue one 
	 denote the monopole BH temperature and  the RN AdS BH one, respectively.}
\label{fig-ads13}
\end{figure}

 Fig.\ref{fig-ads12} confirms this fact. 
 The vertical axis 
shows the discontinuity of BH entropies between the monopole BH at  the cusp point C and 
RN AdS BH. 
It shows how the cusp structure gets small and eventually vanishes at the critical point 
$\tilde \Lambda_{\rm cr (1)}\approx -0.68$.  
We also show how the BH temperatures depend on the cosmological constant 
 in Fig. \ref{fig-ads13}. 

At the critical value of the cosmological constant $\tilde \Lambda_{\rm cr (1)}$,
  these two curves cross. It is consistent with our claim 
   that the phase transition changes from the first 
 order to second order at the critical value  $\tilde{\Lambda}_{\rm cr(1)}$.
  
\begin{figure}
	\includegraphics[width=6cm]{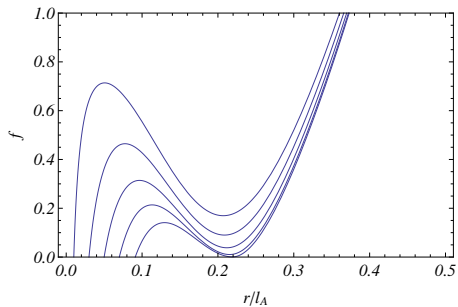}
	\caption{The behavior of $f(\tilde{r})$ fixing $\tilde{\Lambda}=-33$, 
	$\tilde{v}=0.1$ and $\bar{\lambda}=0.1$ but varying $\tilde{r}_{H}$. $ \tilde{r}_{H}=0.01,
	 0.03, 0.05,  0.07$ and $0.091$  from the top. }
\label{fig-ads15}
\end{figure}

In addition, at the second critical value of the cosmological constant 
$\tilde \Lambda_{\rm cr (2)}$, 
another change of the types of phase transition is expected. 
Beyond $\tilde \Lambda_{\rm cr (2)}$, the monopole BH evolves into 
the extreme RN AdS BH as shown in 
Fig. \ref{fig-ads15}.
When the minimum of 
 $f(\tilde{r})$ vanishes as $\tilde r_H$ increases,
  the event horizon of the extreme RN AdS BH 
 appears at the bigger horizon radius ($\ti{r}_{H}\approx 0.220$) 
 than that of the monopole BH
 ($\ti{r}_{H}\approx 0.091$). 
Then the first order phase transition 
 occurs with the discontinuous changes of the BH entropy 
 and the BH temperature. 
  Note that the BH temperatures of the monopole BH at the transition point 
 and of the extreme RN AdS BH are $\tilde{T}\sim O(10^{-5}) \neq 0$ and
  $\tilde{T}=0$, respectively.
The inverse process  from the RN AdS BH to the monopole BH may not occur because the BH entropy must decrease
 at the transition point.

\begin{figure}[h]
	\includegraphics[width=6cm]{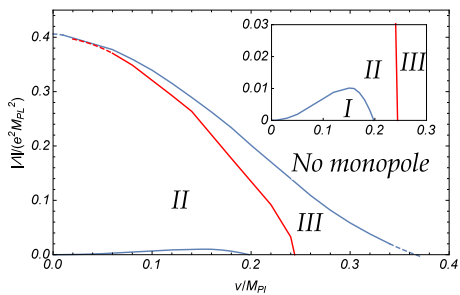}
	\caption{The $v$ dependence of $\Lambda_{\rm cr(1)}, \Lambda_{\rm cr(2)}$
	and $\Lambda_{\rm b}$ for the case of $\bar{\lambda}=0.1$. 
	The curves correspond to the cases of $\Lambda_{\rm cr(1)}$, of
	$\Lambda_{\rm cr(2)}$ and of $\Lambda_{\rm b}$, 
	respectively,  from the bottom. 
	The region I represents where the first order phase transition occurs, 
	while region II denotes where the second order phase transition is expected. 
	In the region III, we find the first order phase transition from 
	the monopole BH to the extreme RN AdS BH.  Beyond $\Lambda_{\rm b}$,
	no monopole BH exists.}
\label{fig-ads14}
\end{figure}

The above two critical values, $\tilde{\Lambda}_{\rm cr(1)}$ and 
 $\tilde{\Lambda}_{\rm cr(2)}$, as well as the boundary value,  
 $\tilde{\Lambda}_{\rm b}$,
depend both on $\tilde{v}$ and $\bar{\lambda}$. 
  Fixing $\bar{\lambda}=0.1$, we show them 
  in Fig. \ref{fig-ads14} .

When we increase the coupling constant $\bar \lambda$, the region I will shrink
and vanish, for example, in the case of $\bar \lambda =1$

\subsubsection{Black Holes in a Heat Bath}
\begin{figure}[h]
	\includegraphics[width=6cm]{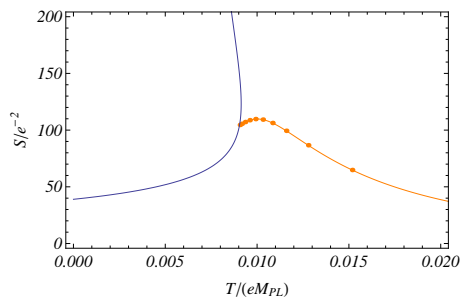}
	\caption{$\tilde{T}$-$\tilde{S}$ relation. 
	($\bar{\lambda}=0.1$, $\tilde{v}=0.1$, and $\tilde{\Lambda}=-0.1$). The blue line represents RN BH and the orange line represent monopole BH. }
\label{fig-ads17}
\end{figure}

\begin{figure}[h]
	\includegraphics[width=6cm]{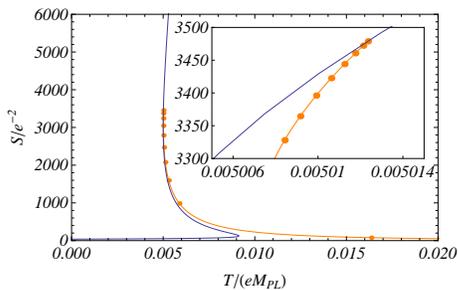}
\caption{$\tilde{T}$-$\tilde{S}$ relation. ($\bar{\lambda}=0.1$, $\tilde{v}=0.001$, and $\tilde{\Lambda}=-1000$). The blue line represents RN BH and the orange line represent monopole BH. }
\label{fig-ads18}
\end{figure}

Next we consider thermodynamical stability of BHs in a heat bath.
We deal with two thermodynamical variables. One is 
the heat capacity, and the other is the Helmholtz free energy.
The heat capacity is defined by 
\begin{equation}
C=T\left(\frac{\partial S}{\partial T}\right).
\label{mono27}
\end{equation}
If the heat capacity is negative, a system in a heat bath 
is  thermodynamically unstable.
While when the system has a positive heat capacity,
such a state is locally stable. 
For the RN AdS BH, its behavior is well studied\cite{BH in ADS}.
As shown  in Figs. \ref{fig-ads17} and \ref{fig-ads18}, 
there exists a finite intermediate range of the entropy, 
where the heat capacity becomes negative. 
Hence the RN AdS BH becomes unstable in such a range.
The unstable BH may change its entropy (the horizon radius)
 via the evaporation or accretion
and evolve into a stable one.
As for the monopole BH branch, the solution appears at 
some point of the RN AdS BH branch.
It may change the evolutional path. 

In the asymptotically flat case, the monopole BH in a heat bath 
is locally unstable for almost all the parameters.
It is because the entropy of the 
monopole BH  usually decreases monotonically
 as the BH temperature increases, giving 
the negative heat capacity.
 However,  when $\bar{\lambda}$ 
is small enough, there exists a thermodynamically stable monopole BH 
in a narrow range of the temperature\cite{monopole BH 5}. 
The parameter region where the stable monopole BHs exist coincides 
with one where the cusp structure appears
 in $\ti{M}$-$\ti{r}_{H}$ relation.
This is because when there exist two monopole BHs with 
the same horizon radius $r_{H}$ (equivalently the same entropy),
the BH entropy must increase in some parameter region.
We find the same feature for the AdS monopole BHs when 
$|\ti{\Lambda}|$ as well as $\bar{\lambda}$ and $\tilde{v}$ are small.

We show some example in Fig. \ref{fig-ads17}.
We choose the parameters $\bar \lambda =0.1, \tilde v=0.1$ and 
$\tilde \Lambda =-0.1$.
The monopole BH branch appears near the marginally stable point of 
RN AdS BHs. 
The heat capacity is first positive near the junction point,
 but it becomes negative soon when the temperature increases.
Hence there exists a stable monopole BH solution for a 
narrow range of the temperature.
This happens when $\ti{\Lambda}>\ti{\Lambda}_{\rm cr(1)}$.

When $\ti{\Lambda}_{\rm cr(1)}>\ti{\Lambda}>\ti{\Lambda}_{\rm b}$,
the entropy of monopole BH decreases monotonically  
in terms of temperature. As a result, there is no 
thermodynamically stable monopole BH. 

In addition, we find 
 new behaviors of thermodynamical stability when $\tilde{v}$ is very small but $|\tilde{\Lambda}|$ is large enough. 
 As shown in Fig. \ref{fig-ads18}, the heat capacity changes its sign 
 through infinity.
 The similar behavior of the stability is found 
 in the case of Einstein-SU(2)-Yang-Mills system with 
 a negative cosmological constant \cite{Thermo SU2YM 2}. 
 This may be explained by the fact  that the monopole BH solution 
 near the RN AdS BH is little affected by the Higgs field.

Next we discuss 
the Helmholtz free energy $F$, which is 
another thermodynamical indicator.
We may judge a global stability by this indicator as we will see.
 Note that the heat capacity is the indicator to judge a
 local stability of a  system in a heat bath.

For convenience, we introduce a dimensionless free energy defined
by 
\begin{equation}
\tilde{F}\equiv e \sqrt{G}F
\,.
\label{mono29}
\end{equation}

In the Einstein-Maxwell theory with a negative cosmological constant,
it is known that if  $|\Lambda|$ is small enough, we find the first order phase transition between the large RN AdS BH and the small RN AdS BH in a thermal bath system \cite{BH in ADS}. The ``swallow tail'' structure appears 
in the $F$-$T$ plane in this phase transition. 
In SO(3)EYMH system with a negative cosmological constant, 
there are three thermal states; RN  AdS BH in a thermal bath, monopole BH in a thermal bath, and a thermal  monopole. 
A thermal  monopole is  a gravitating  monopole in a heat bath 
 with arbitrary temperature, which exists 
as shown in Appendix \ref{HFE}.

\begin{figure}[h]
	\includegraphics[width=6cm]{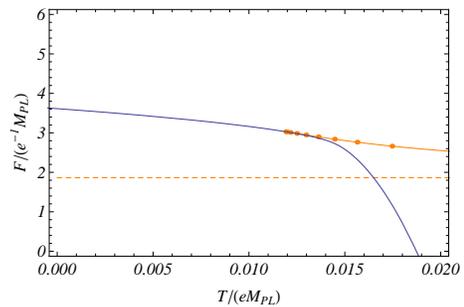}
	\caption{$\tilde{T}$-$\tilde{F}$ relation.($\bar{\lambda}=0.1$, $\tilde{v}=0.1$, and $\tilde{\Lambda}=-1$ ). The blue line represents RN AdS BH and the orange line represents monopole BH. The orange dashed line represents monopole. }
\label{fig-ads19}
\end{figure}

\begin{figure}[h]
	\includegraphics[width=6cm]{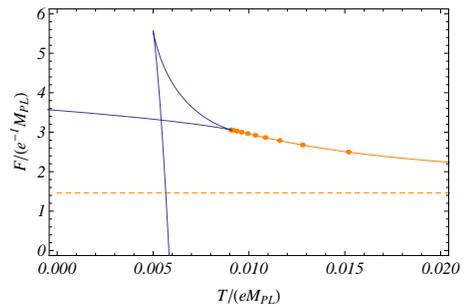}
	\caption{$\tilde{T}$-$\tilde{F}$ relation. 
	($\bar{\lambda}=0.1$, $\tilde{v}=0.1$, and $\tilde{\Lambda}=-0.1$). The blue line represents RN BH and the orange lines represent monopole BH. The orange dashed line represents monopole. }
\label{fig-ads20}
\end{figure}

We show those Helmholtz free energies 
in Figs. \ref{fig-ads19} and \ref{fig-ads20}
for $\tilde{\Lambda}=-1$ and $\tilde{\Lambda}=-0.1$, respectively.
For the large cosmological constant ($\tilde{\Lambda}=-1$),
as shown in Fig. \ref{fig-ads19}, no ``swallow tail'' appears, while
for the small  cosmological constant
 ($\tilde{\Lambda}=-0.1$), 
we find a ``swallow tail'' structure in Fig. \ref{fig-ads20}.
As we have discussed by the heat capacity, there is a thermodynamically stable monopole BH. However it is  just locally stable. 
As seen from Figs. \ref{fig-ads19} and \ref{fig-ads20}, the monopole is favored in the low temperature region while the RN AdS BH becomes mostly stable
 in the high temperature region. 
 The monopole BH is never favored in a thermal bath. 
 As a result, the  Hawking-Page phase transition may occur only between the monopole and the RN AdS BH. 

\section{Concluding Remarks} 
\label{summary}
We have analyzed  the properties of magnetically charged asymptotically  AdS spacetime in 
the EYMH system and its thermodynamical properties. 
Studying the thermodynamical stability, we have discussed the types of phase transition 
for an isolated system and for the system in a heat bath.

The type of phase transition depends on  
$\tilde \Lambda$
as well as  $\tilde v$ and $\bar \lambda$
of the Higgs field. 
In the case of an isolated system,
fixing $\bar \lambda$ small ($\bar \lambda \lsim O(1)$), we find
 two critical values and one boundary value of the cosmological constant, 
 $\tilde \Lambda_{\rm cr (1)}$, 
$\tilde \Lambda_{\rm cr(2)}$,  and $\tilde \Lambda_{b}$,
which depend on $\tilde v$.
The phase diagram on the $\tilde v$-$\tilde \Lambda$
 parameter plane is given by Fig. \ref{fig-ads14}.  
 For the large value of $\bar \lambda\gg O(1)$, 
although the extreme monopole BH was found in the case of $\Lambda=0$
(see the schematic diagram Fig. \ref{phaseD}),
we could not confirm the existence of such a solution because of numerical difficulty.

In the case of the BH system in a heat bath, 
we find the Hawking-Page transition between the RN AdS BH and 
the gravitating monopole. 
The monopole BH does not play anything in this transition, 
although it is locally stable. 

This Hawking-Page transition describes a
 transition between a zero-entropy soliton and a finite-entropy BH spacetime.
Such a phase transition has first been discussed  in the SU(2)-EYM system
\cite{Thermo SU2YM 1}. They showed the possibility of phase transition between Bjoraker-Hosotani soliton and RN AdS BH. The AdS colored BH, which also exists a hairy BH in the system, does not play anything in the transition when the magnetic charge is unity.
Our result of the Hawking-Page phase transition  between AdS monopole and RN AdS BH is similar to this in the SU(2)-EYM system. 
The interesting point is that non-Abelian gauge field leads to the appearance of zero-entropy soliton state. In the SU(2)-EYM system, 
since the magnetic charge changes continuously, 
we can discuss the transition in the large phase diagram.  
The soliton phase smoothly connects to the thermal AdS solution when 
the charge decreases to $0$.
While in our EYMH system, 
the magnetic charge is quantized.
The soliton phase and AdS space are disconnected.
As a result, our phase diagram is restricted.
If we extend the present solution to the dyon, however, 
we may discuss a larger variety of transition as the EYM case.

Since there are many types of phase transitions in the EYMH system, 
it would be interesting to apply this model to the AdS/CFT correspondence.
Since our solution is constructed in four spacetime dimensions, 
we may discuss 3D QFT.
The QFT in 3D flat spacetime corresponding to the 4D EYMH system has already been investigated in \cite{other EYMH 1,other EYMH 3}. The authors solved the Julia-Zee dyon BH solution with asymptotically AdS spacetime and investigated its properties mainly  in the planer limit with/without a backreaction. 
They showed that the non-trivial dyon BH solution is favored at low temperature while the RN dyon BH solution is favored at high temperature,  fixing the chemical potential and the magnetic charge on the boundary, which corresponds to a
 grand canonical ensemble\footnote[3]{Strictly speaking, it is not a ``dyon'' since the magnetic flux on the boundary vanishes in the planer limit.}.
   This phenomenon corresponds to a phase transition between a condensed phase and a normal state of some QFT living in $\mathbb{R}^2 \times \mathbb{R}_{t}^1 $.   
Consideration of the holographic dual of our non-trival BH solution would be interesting because of its variety of phase transition. This work is in progress.

When we are interested in 4D CFT, we have to extend
 our work to the higher dimensions. 
The stable (AdS) gravitating monopole in 4D spacetime is just an extension 
of 't Hooft-Polyakov monopole which is topologically stable in 4D flat space. 
Then first  we must find a 5D topologically stable ``monopole.'' 
However, the known higher dimensional generalization of such a monopole is 
obtained only for an even-spacetime dimensions\cite{high mono}.
As a result, the 5D extension of our study is not straightforward, 
even if it were possible in principle.

\section*{Acknowledgement}
S.M. would like to thank Y. Hoshino and T. Kitamura  for the useful discussion.
This work was supported in part by Grants-in-Aid from the 
Scientific Research Fund of the Japan Society for the Promotion of Science 
(No. 25400276 and No. 16K05362).

\newpage

\appendix
\section{Asymptotically Flat Monopole Black Hole} 
\label{EYMHwocc}

In this Appendix,
 we consider the case without a cosmological constant ($\Lambda=0$) and 
summarize the properties of a monopole black hole in asymptotically flat spacetime.
They were discussed in the pioneering works\cite{monopole BH 1, monopole BH 2, monopole BH 3, monopole stability, monopole BH 4, monopole BH 5, monopole BH 6, monopole BH 7, monopole BH 8, monopole BH 9, monopole BH 10}, finding the dependence of the parameters in the theory, which we show below.

\subsection{Monopole Black Hole Solution}

\begin{figure}[h]
\includegraphics[width=6cm]{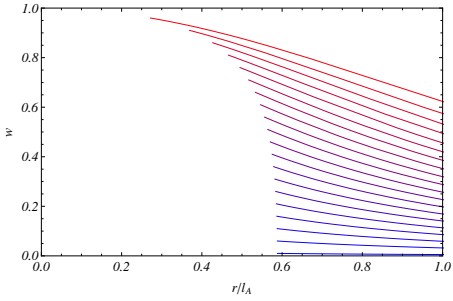}
\caption{The near horizon behavior of  $w(\tilde{r})$ for 
 $\tilde{v}=0.1$ and $\bar{\lambda}=1$.
  Each line denotes the solution with $w_{H}=0.01 \sim 0.96$ at the interval 
  of 0.05 from the bottom curve. }
\label{fig-mono3}
\end{figure}

Fig. \ref{fig-mono3} shows the behavior of the gauge potential $w(\ti{r})$ near the horizon. 
We choose the parameters as $\tilde{v}=0.1$, $\bar{\lambda}=1$.
 $w(\ti{r})$ decreases monotonically and then vanishes 
 at infinity. 
As shown in Fig. \ref{fig-mono3}, the horizon radius $\ti{r}_{H}$ becomes 
larger as $w_{H}$ gets smaller.

\begin{figure}[h]
\includegraphics[width=6cm]{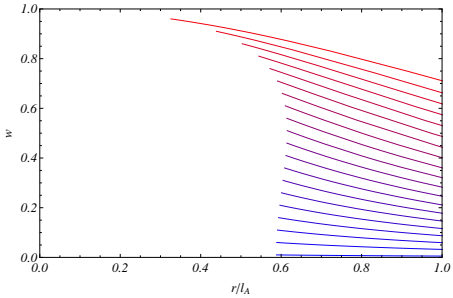}
\caption{The near horizon behavior of  $w(\tilde{r})$ in the case of
 $\bar{\lambda}=0.1$ and $\tilde{v}=0.1$.
  Each line denotes the solution with $w_{H}=0.01 \sim 0.96$ at the interval 
  of 0.05 from the bottom curve. }
\label{fig-mono4}
\end{figure}

However, if $\bar{\lambda}$ is small enough, we find more complicated behavior.
As shown in  Fig.
 \ref{fig-mono4}, when $\bar{\lambda}=0.1$,
there exist two monopole BH solutions with the same horizon radius 
but with the different values of $w_{H}$.

\begin{figure}[h]
	\includegraphics[width=6cm]{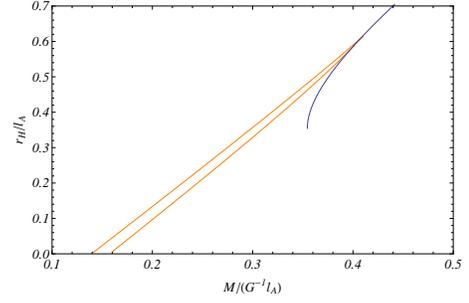}
	\caption{The $\tilde{M}$-$\tilde{r}_{H}$ relations for the cases of 
	 $\bar{\lambda}=1$ (the lower orange curve) and  of $\bar{\lambda}=0.1$
	  (the upper orange curve).
	 We set $\tilde{v}=0.1$. The blue  curve denotes
	 the RN BH  branch.}
\label{fig-mono5}
\end{figure}

\begin{figure}[h]
	\includegraphics[width=7cm]{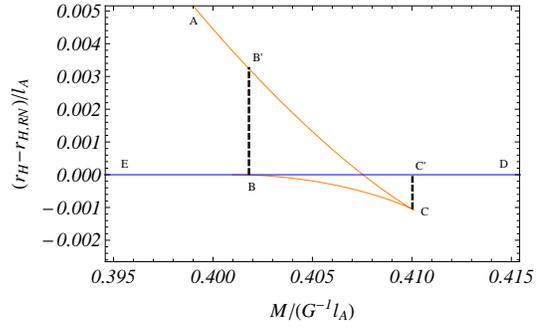}
	\caption{The enlarged figure of Fig.\ref{fig-mono5} near the junction point  
	for the case of 	$\bar{\lambda}=0.1$ and $\tilde{v}=0.1$. 
	The vertical axis denotes the difference of the horizon radii of the monopole BH    
	and 	of the  RN BH. 
	There exist two monopole BH solutions near the junction point B. We  find a 
	cusp structure.
	When the large mass RN BH  reduces its mass via the Hawking radiation, the 
	evolution	path is $ D \to B \to B' \to A$. While, if the small mass 
	monopole BH increases its mass via the matter accretion, 
	the evolution path is $A \to C \to C' \to D$. 
	We expect the first order phase transition in both cases.}
\label{fig-mono7}
\end{figure}

We show the $\tilde{M}$-$\tilde{r}_{H}$ relation in Fig. \ref{fig-mono5}, 
where we plot two cases of $\bar{\lambda}=0.1$ and 1. 
There exist the monopole BH branch and the RN BH one. 
Those two branches connect at one junction point.
Those two figures look similar, but there exists a cusp structure 
near the junction point only for the case of $\bar{\lambda}=0.1$,
which enlarged figure near the junction point 
is shown in Fig. \ref{fig-mono7}. 
There exist two monopole BH solutions near the junction point.
The vertical axis denotes the difference between the horizon radii of the RN BH and 
of the monopole BH.

\begin{figure}[h]
	\includegraphics[width=6cm]{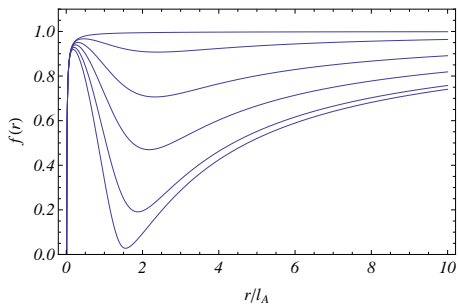}
	\caption{The behaviors of $f(r)=(1-2\tilde{m}(\tilde{r})/\tilde{r}) $ with different values 
	of $\tilde{v}$, i.e., 
	$\tilde{v}=0.01, 0.12, 0.22, 0.3, 0.37, $ and $0.3932\approx \ti{v}_{\rm b}$  
	from the top. 
	We set $ \bar{\lambda}=0, \tilde{r}_{H}=0.01$.}
\label{fig-mono8}
\end{figure}

We should also mention about the existence of the non-trivial solution.
Fig. \ref{fig-mono8} shows the behavior of $f(\tilde{r})$ for a given horizon radius
 $\tilde{r}_{H}$. As $\tilde{v}$ becomes larger, the minimum  value of $f(\tilde{r})$
 decreases and eventually vanishes. For a given value of  $\bar{\lambda}$, there 
 exists a boundary value $\ti{v}_{\rm b}(\bar{\lambda})$, beyond which a
  gravitating monopole solution ceases to exist.
  When $f(\tilde{r})$ vanishes, the zero point becomes 
  a horizon of the extreme RN BH.
  We have only the RN BH solution beyond this critical value $\ti{v}_{\rm b}(\bar{\lambda})$.
  
\subsection{Thermodynamics}
Next we discuss thermodynamical properties of a monopole BH as well as
a trivial RN BH.
We consider two cases; an isolated BH system and BH in a thermal bath.

\subsubsection{Isolated System}
For large $\bar{\lambda}$,  there are only two branches; the RN BH branch and the monopole BH branch (see Fig. \ref{fig-mono5}). 
In an isolated system, the monopole BH branch is thermodynamically favored because 
its entropy is larger than that of the RN BH.
Since there is no monopole BH above the junction point, the RN BH is unique. 

In the present case, the thermodynamical stability is equivalent to 
the dynamical instability.
It was shown that the monopole BH solution is always stable against linear perturbations.
While the RN BH solution becomes unstable below the junction point 
although it is stable above the junction point\cite{monopole BH 3,monopole BH 6}.
We may understand this fact 	intuitively because for such a RN BH,
we have to pack the gauge field and Higgs field inside the horizon 
 radius, which is smaller than the monopole radius $\ell_{A}$. 

This thermodynamical stability indicates the possibility of a phase transition 
between two BHs (the monopole BH and the RN BH).
As we shown in Fig. \ref{fig-mono5}, 
the entropy of the monopole BH with $\bar \lambda=1$ 
increases monotonically 
and connects with the RN BH branch at the junction point.
If matter accretes onto the monopole BH and the mass increases to the junction point,
we expect a second order phase transition from 
the monopole BH to the RN BH.
Conversely, when the BH mass decreases via the Hawking radiation of the RN BH, 
a second order phase transition may occur at the junction point.

The case of small $\bar{\lambda}$ is more interesting. From the $\tilde{M}$-$\tilde{r}_{H}$ relation in Fig.\ref{fig-mono7},  we find three branches; one RN BH branch (the line BD)
and two monopole BH branches (the curves AC and BC).
Using the stability analysis 
 \cite{monopole stability} and the catastrophe theory
 \cite{monopole BH 4,monopole BH 5}, we find that both the  right side RN BH branch 
 of the junction  point B (BD) and the AC monopole branch are locally stable. 
 The BC monopole branch is locally unstable. 
There are two stable states and one unstable state for 
some mass range. We expect the first order phase transition
via the catastrophe theory as follows. 
First we consider the case that matter accretes onto a stable monopole BH. 
Fig. \ref{fig-mono7} shows how such a monopole BH evolves. With mass accretion, the monopole BH evolves along the stable branch AC. When the mass reaches the point C, 
BH jumps onto the point C$'$ on the RN BH branch and 
the  horizon area (the BH entropy)  changes discontinuously.  
It is the first order phase transition.
While, in the case that the BH mass decreases via the Hawking radiation, 
the stable RN BH evolves along the stable branch DB and 
it eventually reaches the point  B. Then it 
 jumps onto the point B$'$ on the stable monopole BH branch.
 
This cusp structure (two monopole branches) vanishes if the coupling constant 
$\bar{\lambda}$ is larger than the critical value $\bar{\lambda}_{\rm cr(1)}(\tilde{v})$,
which depends on the VEV $v$.
Hence 
 the phase transition between the RN BH and the monopole BH is
either

(1) 
the second order when  $\bar{\lambda} > \bar{\lambda}_{\rm cr(1)}$,
\\
or

(2) 
the first order when  $\bar{\lambda} < \bar{\lambda}_{\rm cr(1)}$.

The above story is true only when $\bar \lambda$ and $\tilde{v}$ are rather small
($0\leq \bar \lambda < O(1)$ and $0\leq \tilde{v}\lsim 0.2 $). 

\begin{figure}[h]
	\includegraphics[width=6cm]{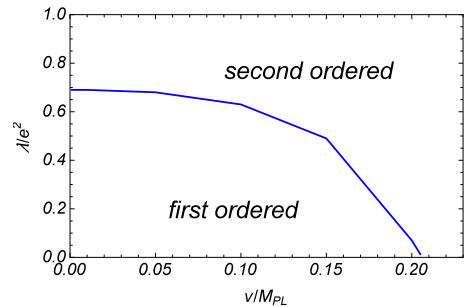}
	\caption{The phase diagram of EYMH system with $\Lambda=0$. Below the blue curve $\bar \lambda=\bar \lambda_{\rm cr (1)}(\tilde{v})$, we find the first order phase transition. While above the critical curve, the second order phase transition is found. }
\label{pd-mono}
\end{figure}

For more general values of the coupling constant $\bar \lambda$ 
and the VEV $\tilde{v}$, we 
may find two more critical curves $\bar \lambda=\bar \lambda_{\rm cr (2)}(\tilde{v})$ 
and $\bar \lambda=\bar \lambda_{\rm b}(\tilde{v})$.
For $\bar \lambda_{\rm cr (1)}(\tilde{v})<\bar \lambda<\bar \lambda_{\rm cr (2)}(\tilde{v})$, we find the second order phase transition as we discussed above.
In the range of  $\bar \lambda_{\rm cr (2)}(\tilde{v})<\bar \lambda<\bar \lambda_{\rm b}(\tilde{v})$, we find a new type of first order phase transition.
There exist the monopole BH branch as well as the RN BH branch, 
but they are not connected. 
The monopole BH transits to the extreme RN BH, at which
 the horizon radius changes discontinuously.
 It is the first order phase transition. 

Beyond $\bar \lambda_{\rm b}(\tilde{v})$, there is no monopole BH.
For a given value of $\bar{\lambda}$, increasing $\tilde{v}$ further,
we find the behavior of the metric function as shown in Fig. \ref{fig-mono8},
i.e.,
 the local minimum of $f(\tilde{r})$ appears,  its value decreases, 
 and eventually vanishes at $\tilde{v}=\tilde{v}_{\rm b}(\bar \lambda)$, 
 beyond which there exists  no monopole BH.  The boundary curve 
 $\tilde{v}=\tilde{v}_{\rm b}(\bar \lambda)$ is equivalent to the curve
   $\bar \lambda=\bar \lambda_{\rm b}(\tilde{v})$.
 
For the very large value of $\bar{\lambda} $,
we may find the similar first order phase transition but 
with the different outcome, that is, 
 the extreme monopole BH  \cite{monopole BH 9}.
 
\begin{figure}[h]
	\includegraphics[width=6cm]{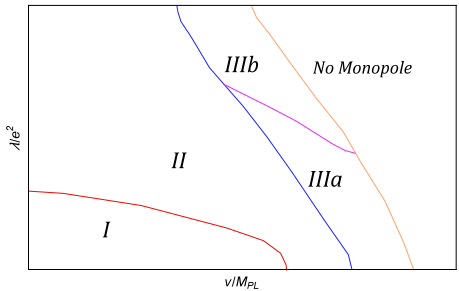}
	\caption{The expected schematic phase diagram in the $\ti{v}$-$\bar \lambda$ plane. The region I represents where the first order phase transition occurs, while region II shows where we find the second order phase transition.
	 Both the region IIIa and IIIb denote where the ``irreversible'' phase transitions appear. 
	 In region IIIa, the phase transition from the monopole BH to the 
	 extreme RN BH occurs, while in the region IIIb, 
	 we expect the transition from the  monopole BH to the extreme monopole BH.}
	\label{phaseD}
\end{figure}

We summarize the expected schematic phase diagram in Fig \ref{phaseD}.

 \subsubsection{BH in a Heat Bath}
\begin{figure}[h]
	\includegraphics[width=6cm]{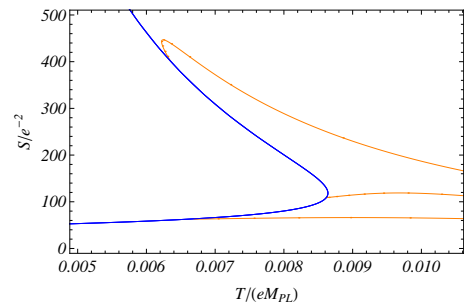}
	\caption{$\tilde{T}$-$\tilde{S}$ relation with varying $\tilde{v}$ ($\bar{\lambda}=0.1$). The blue line represents RN BH and the orange lines represent monopole BH with $\tilde{v}=0.05, 0.1$ and $0.14$  from the top. }
\label{fig-mono10}
\end{figure}

\begin{figure}[h]
	\includegraphics[width=6cm]{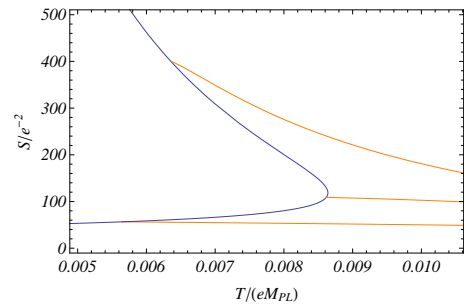}
	\caption{$\tilde{T}$-$\tilde{S}$ relation with varying $\tilde{v}$ ($\bar{\lambda}=1$). The blue line represents RN BH and the orange lines represent monopole BH with $\tilde{v}=0.05, 0.1$ and $0.15$  from the top. }
\label{fig-mono11}
\end{figure}

Next we consider a BH in a heat bath.
In Fig. \ref{fig-mono10}, we show the $\tilde{T}$-$\tilde{S}$ relation 
for  various values of $\tilde{v}$
 fixing $\bar{\lambda}$, which slope describes the heat capacity. 
 If the slope is positive, the system in a heat bath is stable, while if negative, it is 
unstable.
For $\tilde{v}=0.05$, the entropy of the monopole BH changes its slope sign twice;
negative  $\to$    positive $\to$ negative. Hence near the junction point to the 
RH BH branch, 
the system is unstable, and turns to be stable soon, and then becomes unstable.
For $\tilde{v}=0.1$ and larger,
 the sign changes only once; positive $\to$ negative. 
 As a result, the monopole BH in a heat bath is stable near the junction point, 
 but becomes unstable.
For  the small values of $\tilde{v}$, a stable region is very small compared to the
case with the large values of $\tilde{v}$. 
When $\bar{\lambda}$ is large, the behaviors change a lot. 
In Fig. \ref{fig-mono11}, we show  the $\tilde{T}$-$\tilde{S}$ relation for 
$\bar{\lambda}=1$. In this case, the slope of the monopole BH branch is 
always negative, and then there is no stable region.
Only the  RN BH with small entropy is stable.

\begin{figure}[h]
	\includegraphics[width=6cm]{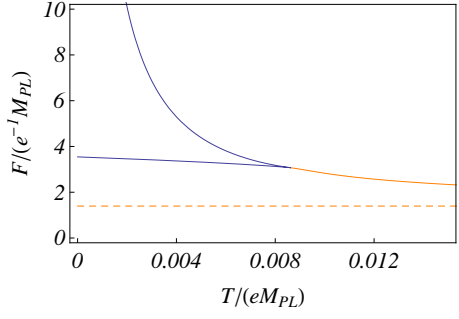}
	\caption{$\ti{T}$-$\ti{F}$relation ($\bar{\lambda}=0.1, \ti{v}=0.1$). The blue line represents RN BH and the orange line represents monopole BH. The orange dashed line represents gravitating monopole.}
\label{free_energy}
\end{figure}

The heat capacity gives an indicator of a thermodynamical stability of a system in 
a heat bath, but it is just a local stability.
In order to discuss the global stability, i.e., which state is mostly preferred, 
we have to analyze the free energy. 
In Fig. \ref{free_energy}, we present the Helmholtz free energy 
of the present system.

The RN BH branch starts from the extreme state with  $T=0$ and turns backward at 
a critical temperature with increasing the free energy. 
In the Einstein-Maxwell system, the first branch which includes the extreme state 
is globally stable. Above the critical temperature,
 no thermal spacetime with magnetic (or electric charge) is possible. In the present EYMH system, however, 
there exist the other branches of the monopole BH and the gravitating monopole. 
Since the free energy of the gravitating monopole is 
always smallest compared with those of the monopole BH as well as of the RN BH,
the gravitating monopole in a heat bath is globally stable.
Hence there is no Hawking-Page transition in  the EYMH system.

\section{Helmholtz Free Energy}
\label{HFE}

In order to discuss a stability of a system in a heat bath,
 we calculate the Helmholtz free energy,
 which is obtained by the partition function defined by the path integral,
  \footnote[4]{This is because we consider only 
  a purely magnetically charged solution.  If we discuss a purely electrically charged solution, 
  the path integral gives the grand partition function\cite{partition}. }

\begin{equation}
Z=\int \mathcal{D}[g,\phi] e^{-I_E[g,\phi]}
\,.
\label{hp1}
\end{equation}

If we assume the 
main contribution in the integration 
comes from the classical solutions of the Euclidean action, 
the partition function is evaluated as 
$Z\approx \exp[-I_{E}^{{\rm on\mathchar`- shell}}]$. 
Then, the Helmholtz free energy $F$ is given by 
\begin{equation}
\beta F=I_{E}^{\rm on\mathchar`- shell}
\label{hp5}
\end{equation}
where $\beta$ is the inverse temperature of the system and 
is identified with 
the Euclidean time period of the on-shell action. 
If we choose an arbitrary value of $\beta$
 as an inverse temperature of the BH spacetime, 
the corresponding Euclidean BH spacetime may have a conical
 singularity at the ``horizon.''  
Since a spacetime with a conical singularity 
 does not contribute to the partition function, 
the only allowed value of $1/\beta$ must be the Hawking temperature
$T$ given by Eq. (\ref{mono24}). 

On the other hand, in the case of a gravitating monopole, 
an arbitrary period of the Euclidean time  $\beta$ is possible.
We find a thermal gravitating soliton spacetime with arbitrary temperature $T$.

Denoting a spacetime by $ \mathcal{M }$ with the metric $g_{\mu\nu}$ and its boundary 
by $ \partial \mathcal{M} $ with the induced metric $h_{mn}$, 
the action of the gravity system is given by
\begin{equation}
I=I_{\rm EH}+I_{\rm GHY}+I_{\rm C}+I_{\rm matter}
\,,
\end{equation}
where
\begin{eqnarray}
\displaystyle I_{\rm EH}&=&\frac{1}{16 \pi G} \int_{\mathcal{M}} d^{4}x \sqrt{-g}(R-2\Lambda  ) 
\\
\displaystyle I_{\rm GHY} &=& \frac{1}{8 \pi G} \int_{\partial \mathcal{M}} d^{3} x \sqrt{|h|} \varepsilon K \\
\displaystyle I_{\rm C} &=& \frac{1}{8 \pi G} \int_{\partial \mathcal{M}} d^{3} x \sqrt{|h|} 
\varepsilon K_{0} \\
\end{eqnarray}
where $ \varepsilon = g(\vect{n}, \vect{n}) $ with 
$ \vect{n}= n^{\mu} \partial_{\mu} $ being a normal vector of a hypersurface, 
$K=h^{ab}K_{ab}$ is the trace of an extrinsic curvature 
$K_{ab}=\frac{1}{2}(\nabla_{\mu} n_{\nu}+\nabla_{\nu} n_{\mu})e^{\mu}_a e^{\nu}_b$, 
and $K_{0} $ is the counter term to remove the divergence.
Since the Einstein-Hilbert term $I_{\rm EH}$ contains the second derivative of the metric, 
we have to introduce the Gibbons-Hawking-York term $I_{\rm GHY}$ in order to write the
 entire action only with the metric and the first derivative of the metric. The third term 
 $I_{\rm C}$
  is the so-called counter term 
  and introduced to remove the divergence appeared in the Einstein-Hilbert term and the Gibbons-Hawking-York term. 
  
  In our setup, we take $K_{0}=
   -\frac{2}{L}\left(1+\frac{L^2}{2}R^{(3)} \right)$ where $R^{(3)}$ is 
   the Ricci scalar    on the boundary, which is given by  $R^{(3)}=2/r^2$
 for a  static and spherically symmetric spacetime, where 
   $L \equiv \sqrt{-3/\Lambda}$ is AdS radius. 
   
 Substituting our ansatz of the solution, the Euclideanized on-shell action 
 $I_E^{\rm on\mathchar`- shell}$ 
   is  written

\begin{widetext}
\begin{equation}
\begin{array}{l}
I_{E}^{\rm on\mathchar`- shell}=I_{E,{\rm EH}}^{\rm on\mathchar`- shell}
+I_{E,{\rm matter}}^{\rm on\mathchar`- shell}+I_{E,{\rm GHY}}^{\rm on\mathchar`- shell}
+I_{E,{\rm C}}^{\rm on\mathchar`- shell} \\
 \\
\displaystyle I_{E,{\rm EH}}^{\rm on\mathchar`- shell}+I_{E,{\rm matter}}^{\rm on\mathchar`- shell}=\displaystyle 
-4\pi \int_{0}^{\beta} dt_{E} \int_{\rm bulk} dr \ e^{-\delta} r^{2} \left[ 
\frac{\Lambda}{8\pi G} -f \frac{(w')^{2}}{e^{2}r^{2}}-\frac{1}{2}
\frac{(w^{2}-1)^{2}}{e^{2}r^{4}}+\frac{\lambda}{4}v^{4}(h^{2}-1)^{2} \right]
\,,\\
 \\
\displaystyle I_{E,{\rm GHY}}^{\rm on\mathchar`- shell}= \displaystyle \left. -\frac{1}{2G}\int_{0}^{\beta} 
dt_{E} \ e^{-\delta} r^{2}\left[ \left( \frac{1}{2}f'-f\delta' \right)+\frac{2}{r}f  
\right] \right|_{r \to \infty}\,,\\
 \\
\displaystyle \left. I_{E,{\rm C}}^{\rm on\mathchar`- shell}= \displaystyle \frac{1}{2G}\int_{0}^{\beta} dt_{E} 
\ e^{-\delta} r^{2}\sqrt{f}\frac{2}{L}\left( 1+\frac{1}{2}\frac{L^{2}}{r^{2}} \right) 
\right|_{r\to\infty} 
\,,
\\
\end{array}
\label{hp6}
\end{equation}
\end{widetext}
where
$t_{E}$ is the Euclidean time with the period of $\beta$.\\
 The integration bulk region depends on whether the state
 is the BH solution or the soliton solution; in the case of the BH solution,
  the region is $[r_{H}, \infty)$, while  it is $[0, \infty)$
for the soliton solution.

The Helmholtz free energy is the indicator
 to know the most preferred state in a thermal bath.
If there exist multiple BH solutions with the same temperature,
 the minimum of the free energy
chooses the most favored state  in a thermal bath system. 
Such a state is globally stable in a thermal bath.

Once we know the solution of the EYMH system, we can evaluate 
$I_{E}^{\rm on\mathchar`- shell}$ by Eq. (\ref{hp6}),
and then find the Helmholtz free energy $F$, which is given by Eq. (\ref{hp5}).


\end{document}